\documentclass[journal,twoside,web]{ieeecolor}
\usepackage{generic}
\usepackage{cite}
\usepackage{amsmath,amssymb,amsfonts}
\usepackage{algorithmic}
\usepackage{graphicx}
\usepackage{algorithm,algorithmic}
\usepackage{newunicodechar}
\newunicodechar{₀}{$_0$}
\usepackage[
  colorlinks=false, 
  pdfborder={0 0 1}, 
  linkbordercolor={1 0 0}, 
  citebordercolor={1 0 0}, 
  urlbordercolor={1 0 0},  
  hypertexnames=false
]{hyperref}
\hypersetup{hidelinks=true}
\usepackage{textcomp}
\usepackage{array}
\usepackage[caption=false,font=footnotesize,labelfont=rm,textfont=rm]{subfig}
\usepackage{stfloats}
\usepackage{url}
\usepackage{verbatim}
\usepackage{booktabs}
\usepackage{multirow}
\usepackage{cite}
\usepackage{threeparttable}

\newcommand{\smallpm}[1]{\raisebox{-0.4ex}{\scriptsize $\pm$#1}}

\def\BibTeX{{\rm B\kern-.05em{\sc i\kern-.025em b}\kern-.08em
    T\kern-.1667em\lower.7ex\hbox{E}\kern-.125emX}}
\begin{document}
\title{Towards Robust Assessment of Pathological Voices via Combined Low-Level Descriptors and Foundation Model Representations}

\author{Whenty Ariyanti, \IEEEmembership{Student Member, IEEE}, Kuan-Yu Chen, \IEEEmembership{Member, IEEE}, Sabato Marco Siniscalchi, \IEEEmembership{Member, IEEE}, Hsin-Min Wang, \IEEEmembership{Senior Member, IEEE}, and Yu Tsao, \IEEEmembership{Senior Member, IEEE}
\thanks{This work was supported in part by the National Science and Technology Council and Academia Sinica. (Corresponding author: Yu Tsao.)}
\thanks{Whenty Ariyanti is with the Department of Computer Science and Information Engineering, National Taiwan University of Science and Technology, Taipei 106, Taiwan, and also with the Research Center for Information Technology Innovation, Academia Sinica, Taipei 11529, Taiwan (e-mail: d11115805@mail.ntust.edu.tw).}
\thanks{Kuan-Yu Chen is with the Department of Computer Science and Information Engineering, National Taiwan University of Science and Technology, Taipei 106, Taiwan (e-mail: kychen@mail.ntust.edu.tw).}
\thanks{Sabato Marco Siniscalchi is the University of Palermo, Palermo, Italy (e-mail: sabatomarco.siniscalchi@unipa.it). }
\thanks{Hsin-Min Wang is with the Institute of Information Science, Academia Sinica, Taipei 11529, Taiwan (e-mail: whm@iis.sinica.edu.tw).}
\thanks{Yu Tsao is with the Research Center for Information Technology Innovation, Academia Sinica, Taipei 11529, Taiwan, and also with the Department of Electrical Engineering, Chung Yuan Christian University, 11529, Taiwan (e-mail: yu.tsao@citi.sinica.edu.tw).}}

\maketitle

\begin{abstract}
Perceptual voice quality assessment plays a vital role in diagnosing and monitoring voice disorders. Traditional methods, such as the Consensus Auditory-Perceptual Evaluation of Voice (CAPE-V) and the Grade, Roughness, Breathiness, Asthenia, and Strain (GRBAS) scales, rely on expert raters and are prone to inter-rater variability, emphasizing the need for objective solutions. This study introduces the Voice Quality Assessment Network (VOQANet), a deep learning framework that employs an attention mechanism and Speech Foundation Model (SFM) embeddings to extract high-level features. To further enhance performance, we propose VOQANet+, which integrates self-supervised SFM embeddings with low-level acoustic descriptors—namely jitter, shimmer, and harmonics-to-noise ratio (HNR). Unlike previous approaches that focus solely on vowel-based phonation (PVQD-A), our models are evaluated on both vowel-level and sentence-level speech (PVQD-S) to assess generalizability. Experimental results demonstrate that sentence-based inputs yield higher accuracy, particularly at the patient level. Overall, VOQANet consistently outperforms baseline models in terms of root mean squared error (RMSE) and Pearson correlation coefficient across CAPE-V and GRBAS dimensions, with VOQANet+ achieving even greater performance gains. Additionally, VOQANet+ maintains consistent performance under noisy conditions, suggesting enhanced robustness for real-world and telehealth applications. This work highlights the value of combining SFM embeddings with low-level features for accurate and robust pathological voice assessment.
\end{abstract}

\begin{IEEEkeywords}
Pathological voice quality assessment, VOQANet, CAPE-V, GRBAS, speech foundation models, voice disorder assessment
\end{IEEEkeywords}

\section{Introduction}
\label{sec:introduction}
\IEEEPARstart{V}{oice} disorders are common in modern society, and pathological voice quality can seriously affect an individual's communication ability and social well-being \cite{Syusyiang2022}, \cite{Ariyanti2021}. They arise from various conditions, including vocal fold nodules, polyps, paralysis, neurological diseases, such as Parkinson's disease, and head and neck cancers \cite{pvqd}. These disorders affect vocal characteristics, such as hoarseness, breathiness, roughness, or strain, requiring perceptual examination by trained clinicians. Therefore, vocal signal analysis has become a widely used non-invasive screening tool in otolaryngology, neurology, and speech-language pathology clinics \cite{kreiman,Patel2018}.
Voice Quality Assessment (VQA) aims to improve the diagnosis, monitoring, and treatment of voice disorders by providing an objective and standardized assessment of vocal function. It serves as a critical tool for identifying pathological voice conditions, tracking disease progression, and evaluating the effectiveness of therapeutic interventions. 
Traditionally, VQA relies on perceptual assessments by experienced clinicians using standardized scales, such as the Consensus Auditory-Perceptual Evaluation of Voice (CAPE-V) and the Grade, Roughness, Breathiness, Asthenia, Strain (GRBAS) \cite{CapeV},\cite{grbas},\cite{sasougrbas}. CAPE-V has been adapted to multiple languages, including French, Turkish, European Portuguese, and Japanese, further demonstrating its clinical relevance and international applicability~\cite{CAPEV-French}, \cite{CAPEV-Turkish},\cite{jesus}, \cite{CAPEV-Japanese}. It provides continuous ratings (0–100) for perceptual attributes, whereas GRBAS uses a discrete 4-point ordinal scale (0–3). Recently, CAPE-V has been revised to CAPE-Vr (Consensus Auditory-Perceptual Evaluation of Voice—Revised) \cite{CAPE-Vr}, with updated recommendations for clinical use. It has been noted that clinicians experienced challenges in rating attributes such as overall severity, strain, and pitch when using the original CAPE-V, highlighting the need for clearer definitions and more consistent administration. 
CAPE-Vr addresses these issues through a revised rating form, updated stimuli, and expanded categories, thereby providing more precise guidance for clinical assessment while maintaining the intent of the original protocol.
For both, the larger the value, the more severe the condition. Although these perceptual ratings are widely used and provide valuable qualitative insights into voice disorders \cite{{Hirano1981}}, they are inherently subjective and prone to inter- and intra-examiner variability. Recent studies have explored other strategies, such as crowdsourcing perceptual ratings, particularly in neurological disorders like Parkinson’s disease, to improve scalability and maintain rating validity~\cite{McAllister2023}. The reliance on expert raters makes standardization difficult and increases the need for automated VQA. 

Machine learning (ML) and deep learning approaches have been explored to address these challenges \cite{grbasxie}. Traditional ML models, such as Random Forest (RF), Support Vector Machine (SVM), and k-Nearest Neighbors (KNN), have been used to predict CAPE-V scores based on low-level speech descriptors (LLDs) \cite{TFMBased},\cite{LightWeight}, but struggle to capture the complexity of pathological voices. To overcome these limitations, recent studies have employed deep learning and ensemble frameworks for pathological voice classification~\cite{Ariyanti2021}, demonstrating the benefits of combining multiple modalities and learned representations for more robust voice assessment. To further enhance clinical VQA, attention-based models have also been proposed for predicting GRB scores from sustained vowel phonation, showing improved accuracy over earlier neural architecture \cite{Hann2023}. Recent advances in Speech Foundation Models (SFMs), including WavLM \cite{WavLM}, HuBERT \cite{HuBERT}, and Whisper \cite{Whisper}, have performed well in extracting high-level speech representations.
WavLM and HuBERT are trained using self-supervised learning (SSL), while Whisper adopts a semi-supervised paradigm, leveraging large-scale audio-text pairing data collected through weak supervision. Beyond their success in general speech processing, HuBERT has been shown to be particularly effective in detecting pathological voices~\cite{koudounas2024}. Whisper has also been explored in clinical speech-language applications, including post-stroke speech and language assessment~\cite{davudova2025}. Furthermore, WavLM has been employed for pathological voice analysis in systematic reviews and applied studies~\cite{sindhu2024}.
These models provide robust and transferable features, which are well-suited for downstream tasks, including VQA. WavLM’s denoising pre-training and Whisper’s multilingual coverage further motivate their use in VQA.

In this study, we propose VOQANet (Voice Quality Assessment Network), a deep learning-based framework with an attention mechanism that leverages SFM embeddings for perceptual VQA. We further introduce VOQANet+, which integrates LLDs such as jitter, shimmer, and harmonics-to-noise ratio (HNR) \cite{Teixeira2013} with SFM embeddings to enhance robustness and clinical interpretability. This combination enables VOQANet+ to benefit from both high-level learned representations and complementary low-level signal-based features. 
Both models are evaluated on the Perceptual Voice Quality Dataset (PVQD), reporting both utterance-level and patient-level results, where the latter predictions are averaged across each speaker's utterances. Experimental results show that VOQANet provides a strong baseline, while VOQANet+ consistently improves prediction accuracy and generalization, especially under noisy conditions. 

The main contributions of this study are summarized as follows: First, we propose VOQANet, a deep learning framework with an attention mechanism that systematically evaluates the effectiveness of SFM embeddings of pre-trained speech models for perceptual VQA. Second, we propose VOQANet+, an extended version of VOQANet that combines LLDs with SFM embeddings to improve model interpretability and performance by integrating domain-specific knowledge. Third, we conduct a comprehensive evaluation on the PVQD dataset, including both utterance-level and patient-level evaluations, aligning with clinical assessment practice. Through comprehensive evaluations on the PVQD dataset, our results demonstrate the potential of SFM-driven methods for robust, interpretable, and clinically relevant automated VQA.

\section{Related Work}
\subsection{Automated Pathological Voice Quality Assessment}
Ensuring robustness and generalizability is critical for real-world applications of perceptual VQA. Traditional methods rely on ML models, such as RF, SVM, and KNN using LLDs, like jitter, shimmer, zero crossing rate, and HNR \cite{LightWeight}. A lightweight feature extraction method has been proposed to leverage these models for CAPE-V prediction. However, despite offering interpretability and domain relevance, such models often struggle to generalize across datasets due to speech variability and the sensitivity of LLDs to recording conditions~\cite{LightWeight}. While CAPE-V provides continuous ratings and GRBAS uses a discrete scale, both are susceptible to inter-rater differences~\cite{kreiman2007}, \cite{nagle}, further motivating objective and automated assessment. 

To overcome these limitations, recent work has adopted deep learning-based methods, especially leveraging SFMs such as Whisper \cite{Whisper}, and WavLM \cite{WavLM}, which learn rich acoustic representations from large-scale raw waveforms. WavLM includes a denoising pre-training objective that improves robustness to background noise and acoustic variability. This feature is particularly beneficial for disordered speech, which often deviates from typical acoustic patterns. On the other hand, Whisper is trained on a large-scale multilingual and multitask corpus, achieves strong performance in ASR, and has been explored in tasks such as speaker verification (SV)~\cite{Liu2024}. These findings highlight the potential of SFMs in clinical applications but also underscore that embeddings learned from general speech may not fully capture clinically salient characteristics relevant to voice pathology. To overcome this limitation, hybrid approaches that combine SFM embeddings with LLDs have been explored, showing improved robustness and interpretability in demanding speech tasks~\cite{Elbanna2022}. Based on these insights, we explore deep learning-based methods that combine SFM representations with clinically relevant feature representations.

\subsection{Speech Foundation Models}
Recent advances in SSL have introduced SFMs, which provide more robust and generalizable representations for downstream speech tasks. These models have attracted much attention in the speech processing community due to their ability to learn meaningful representations directly from raw audio.
Models such as WavLM and HuBERT are trained using SSL, learning contextual representations from unlabeled data, whereas Whisper adopts a semi-supervised approach leveraging large-scale audio–text pairs collected with weak supervision. Compared to traditional supervised methods, SFMs can capture low-level acoustic features and high-level linguistic patterns without the need for extensive annotations.

Models pre-trained on large-scale corpora such as Librispeech~\cite{panayotov2015librispeech} have achieved excellent performance in various tasks such as ASR, SV, speech synthesis, and speech emotion recognition \cite{SSLSurvey,pepino2021comparison}. 
In VQA, SFMs have been explored for their ability to extract rich, contextualized acoustic representations linked to perceptual attributes such as breathiness, strain, and roughness. These deep representations are able to capture complex speech patterns that are usually difficult to model using only LLDs. 

\subsection{Hybrid Models Combining SFM and LLDs}
Combining SFM embeddings with LLDs has been explored as a way to capture both data-driven and clinically interpretable speech characteristics. A hybrid approach that integrates a BYOL-derived model with LLDs extracted using openSMILE has shown strong performance in speech analysis tasks \cite{Elbanna}. Similarly, self-supervised learning (SSL)–based speech models have outperformed traditional acoustic features and physiological parameters such as heart-rate–related measures, further demonstrating the effectiveness of combining SSL embeddings with LLDs \cite{Elbanna2022}. In \cite{Hung} and \cite{zezario2023mosa}, the combined features derived from SSL-based representations and LLDs achieved improved performance in both speech enhancement and speech assessment tasks, respectively, compared with using either feature type alone. In this work, we explore, for the first time, a hybrid design for clinical pathological voice assessment, targeting both GRBAS and CAPE-V prediction. Although prior clinical findings suggest that CAPE-V ratings may correlate more strongly with objective acoustic and aerodynamic measures than GRBAS in certain populations~\cite{Fujiki2021}, both scales remain widely used in perceptual voice quality assessment and provide complementary insights. Therefore, we include both GRBAS and CAPE-V in this study.

\subsection{Evaluation Strategies in Voice Assessment Models}
Assessment strategies play a critical role in evaluating the reliability and clinical applicability of voice assessment models. Most prior studies on automated VQA focus on predicting perceptual scores at the utterance level, treating each audio segment as an independent sample~\cite{LightWeight,Elbanna}. While this approach enables fine-grained analysis, it may not capture broader patterns of the patient's overall vocal characteristics, especially when phonetic content varies across utterances.
To enhance clinical reliability, recent studies have explored patient-level assessments by aggregating predictions across multiple utterances from the same speaker~\cite{Elbanna2022}. This strategy better reflects the real-world diagnostic process, where clinicians assess pathological voice quality based on a complete set of speech samples rather than isolated segments.

\begin{figure*}[t]\centering
  \centering
  \centerline{\includegraphics[width=0.98\textwidth]{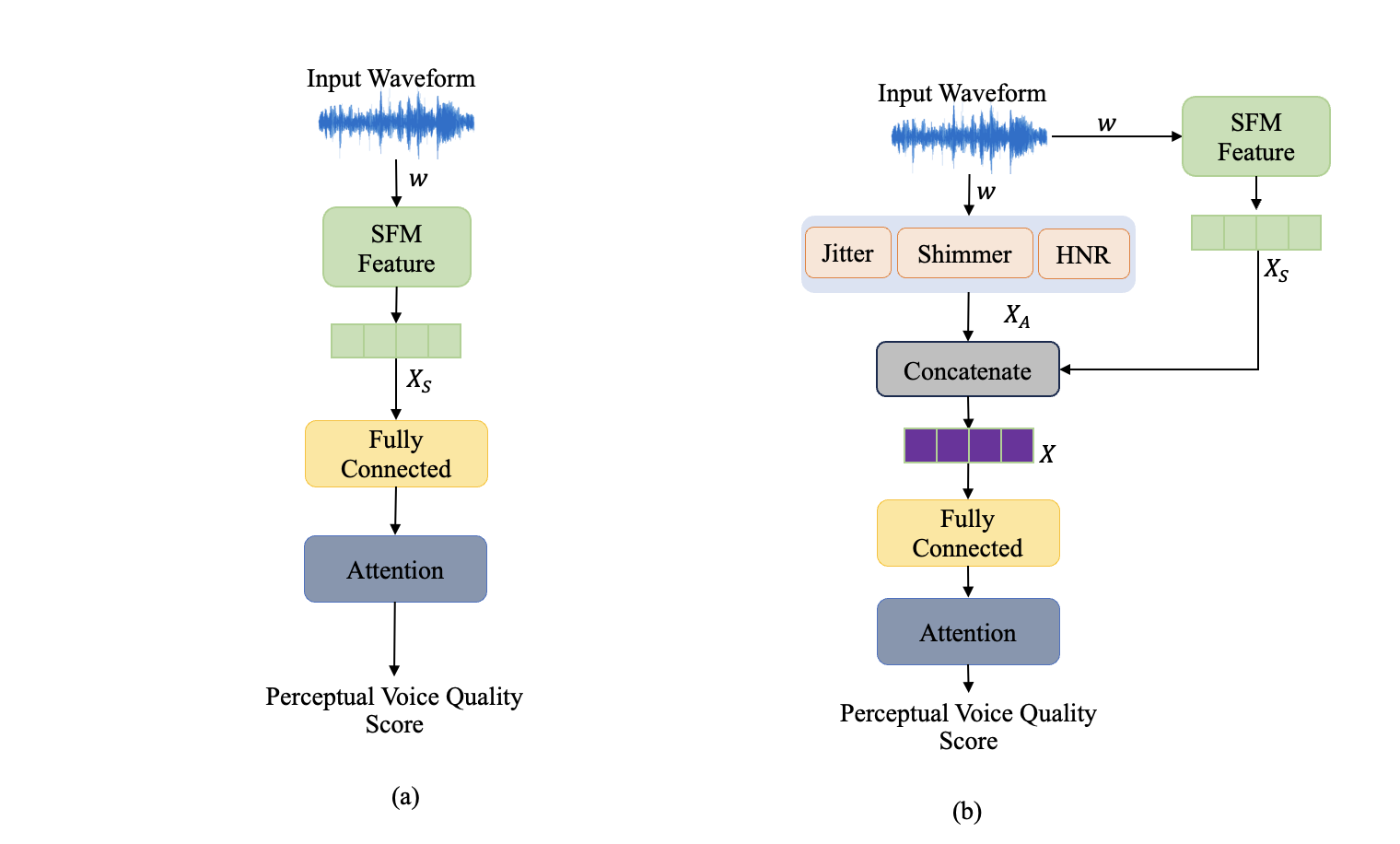}}
\caption{Architectures of VOQANet (a) and VOQANet+ (b).}

\label{modelarc}
\end{figure*}

\begin{figure*}[t]\centering
  \centering
  \centerline{\includegraphics[width=0.98\textwidth]{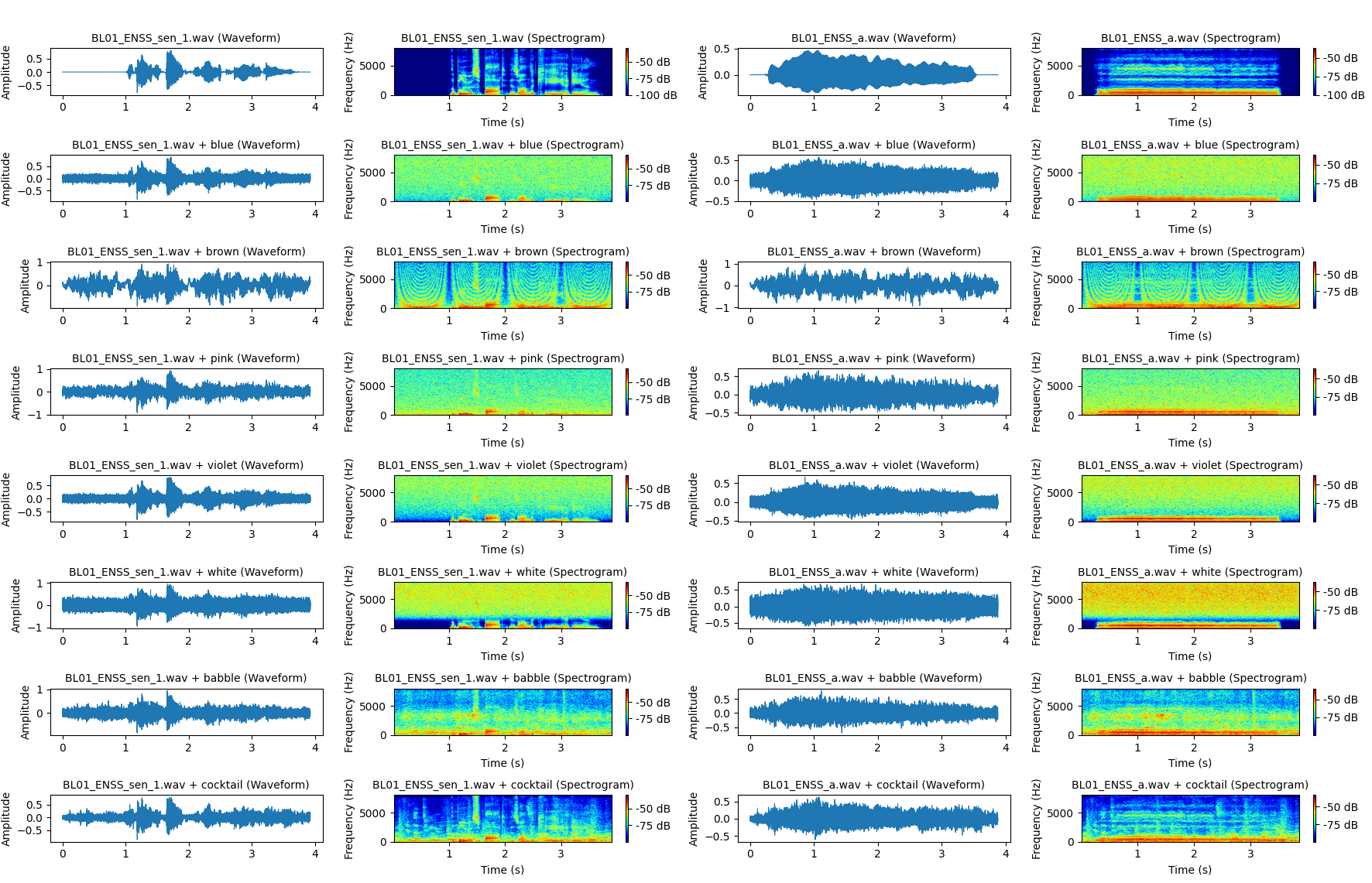}}
\caption{Waveforms and Spectrograms of audio samples in PVQD-A and PVQD-S}

\label{waveformspectrogram}
\end{figure*}

\section{Proposed Method}
The overall architectures of VOQANet and VOQANet+ are shown in Fig.~\ref{modelarc}. VOQANet only uses SFM embeddings as input, while VOQANet+ combines SFM and LLDs as input.

\subsection{VOQANet: SFM-Based Feature Learning}
As shown in Fig.~\ref{modelarc}(a), VOQANet leverages SFM embeddings extracted by a pre-trained model to capture rich acoustic and prosodic characteristics of the input waveform. Given a waveform $\textit{\textbf{w}}$, the SFM embedding is calculated as a weighted sum of the hidden representations of all transformer layers:
\begin{equation}
\boldsymbol{X}_{S} = \sum_{\ell=0}^{L} \alpha_\ell \cdot \mathbf{h}^{(\ell)}(\textit{\textbf{w}}),
\end{equation}
where $\mathbf{h}^{(\ell)}$ represents the hidden state at layer $\ell$ of the pre-trained SFM, and $\alpha_\ell$ is a learnable scalar weight normalized by softmax. This layer-wise aggregation strategy follows prior work that showed that weighted combinations of intermediate layers outperform single-layer embeddings in speech assessment tasks~\cite{chianghasa}. This mechanism allows the model to adaptively emphasize the layers most relevant to pathological voice quality.

\subsection{VOQANet+: Joint Representation Learning}
As shown in Fig.~\ref{modelarc}(b), to further enhance the model’s ability to capture clinically relevant speech characteristics, VOQANet+ combines LLDs with SFM embeddings. These LLDs include Jitter, Shimmer, and HNR, which are extracted from the same waveform $\textit{\textbf{w}}$ using signal processing techniques:
\begin{equation}
\boldsymbol{X}_{A} = f_{A}(\textit{\textbf{w}}),
\end{equation}
where $f_{A}(\cdot)$ denotes the LLD feature extraction function. The final feature representation is obtained by concatenating the two feature types:
\begin{equation}
\boldsymbol{X = [X}_{S}^{\mathsf{T}}| \boldsymbol{X}_{A}^{\mathsf{T}}]^{\mathsf{T}}.
\end{equation}
As a result, a 1027-dimensional hybrid feature vector sequence is obtained. This combination enables the model to jointly learn from high-level SFM embeddings and LLDs, thereby improving the robustness and interpretability of perceptual VQA.

\subsection{Model Architecture and Training}
Both VOQANet and VOQANet+ use the same regression backbone: a three-layer fully connected neural network with batch normalization, dropout, and ReLU activation:
\begin{equation}
\textbf{H} = \sigma(\mathbf{W}_3 \cdot (\sigma(\mathbf{W}_2 \cdot (\sigma(\mathbf{W}_1 \boldsymbol{X} + \mathbf{b}_1)) + \mathbf{b}_2)) + \mathbf{b}_3),
\end{equation}
where $\mathbf{H}$ is the latent representation, and $\sigma(\cdot)$ denotes the ReLU function. To make the network focus on the salient components of the learned features, attention-based pooling is applied after the last hidden layer to calculate a weighted summary of the feature representation. The attention module first projects $\mathbf{H}$ into an intermediate space using a non-linear transformation, and then calculates the attention weights over the feature dimension:
\begin{equation}
\alpha = \text{softmax}(\mathbf{W}_{\text{attn}} \tanh(\mathbf{W}_h \mathbf{H} + \mathbf{b}_h) + \mathbf{b}_{\text{attn}}).
\end{equation}
where $\mathbf{W}_h$ and $\mathbf{b}_h$ are the learnable parameters of the hidden projection that map $\mathbf{H}$ into an attention space, and $\mathbf{W}_{\text{attn}}$ and $\mathbf{b}_{\text{attn}}$ are the parameters of the output projection used to compute the attention logit. The attended feature vector $\textit{\textbf{z}}$ is obtained as follows,  
\begin{equation}
\textit{\textbf{z}} = \sum_{t=1}^{T} \alpha_t \cdot \mathbf{H}_t,
\end{equation}
where $T$ is the length of the feature vector sequence, and $\mathbf{H}_t$ represents the feature vector at position $t$ in the sequence.
Finally, $\textit{\textbf{z}}$ is fed into the final regression layer to predict the VQA score. 

To prioritize clinically significant deviations, we use the Weighted Mean Squared Error (WMSE) loss function~\cite{lu2021perceptually}:
\begin{equation}
\mathcal{L}_\text{WMSE} = \frac{1}{N} \sum_{i=1}^{N} \left(1 + \frac{Y^i}{Y_{\max}} \cdot \beta\right) \cdot (\hat{Y}^i - Y^i)^2,
\end{equation}
where $N$ is the number of training samples; $\beta$ is a hyperparameter that controls the degree of emphasis on higher severity levels; $Y^i$ and $\hat{Y}^i$ are the predicted and ground-truth ratings of sample $i$, respectively; and $Y_{\max}$ is the maximum ground-truth score among the $N$ training samples.



\section{Experiments}
\subsection{Dataset}\label{AA}
The audio samples used in this study come from the Perceptual Voice Quality Database (PVQD) provided by The Voice Foundation~\cite{pvqd}. The dataset includes 296 recordings, each containing sustained /a/ and /i/ vowels and connected-speech samples following the six standardized CAPE-V sentences~\cite{CapeV}. These sentences are part of the official CAPE-V protocol and do not involve number counting or spontaneous speech, designed to elicit diverse phonetic contexts and form part of the official CAPE-V protocol, ensuring the material reflects clinically relevant speech patterns. Combining sustained vowels with CAPE-V sentences aligns with standard dysphonia assessment practices, capturing both glottal-source characteristics and connected-speech dynamics. All recordings are stored in 16-bit WAV format at a 44.1 kHz sampling rate. In addition to raw audio, the dataset also provides metadata on age, gender, diagnosis, and expert perceptual ratings for both CAPE-V and GRBAS. CAPE-V scores are continuous, ranging from 0 to 100, and are used to assess dimensions such as overall severity, breathiness, and strain. Although CAPE-V has been revised to CAPE-Vr, the PVQD corpus provides annotations based on the original CAPE-V framework, so these CAPE-V annotations were adopted in this study. GRBAS scores use a 0–3 ordinal scale~\cite{Kojima2024} and are provided for both vowel (PVQD-A) and connected-speech (PVQD-S) samples, enabling direct comparison across phonation tasks. This represents a data-driven extension of the GRBAS framework beyond its traditional clinical use. Each recording was independently rated by two qualified speech-language pathologists, and the final perceptual score for each dimension is the average of the two raters to ensure reliability and reduce subjective bias. Table~\ref{tab:capev_grbas_stats} summarizes the score distribution of each dimension of the two scales.

\begin{table}[!t]
    \caption{Descriptive Statistics of CAPE-V and GRBAS Ratings in the PVQD Dataset (N=296)}.
    \begin{tabular}{llccccc}
        \toprule
         Scale & Attribute & Mean & Median & Mode & Min & Max \\
         \midrule
         \multirow{6}{*}{\shortstack{CAPE-V\\(0--100)}}

            & Severity     & 29.4  & 19.5 & 19.3 & 0.33 & 98.67 \\
            & Roughness    & 20.7  & 13.7 & 9.7  & 0.17 & 84.83 \\
            & Breathiness  & 19.8  & 12.2 & 5.0  & 0.00 & 99.50 \\
            & Strain       & 21.1  & 12.2 & 4.5  & 0.12 & 96.83 \\
            & Pitch        & 16.3  & 9.3  & 0.5  & 0.00 & 99.17 \\
            & Loudness     & 18.7  & 8.8  & 0.7  & 0.00 & 99.17 \\
        \midrule
        \multirow{5}{*}{\shortstack{GRBAS\\(0--3)}}

            & Grade        & 1.0   & 0.8  & 0    & 0    & 3 \\
            & Roughness    & 0.8   & 0.7  & 0    & 0    & 3 \\
            & Breathiness  & 0.7   & 0.4  & 0    & 0    & 3 \\
            & Asthenia     & 0.6   & 0.2  & 0    & 0    & 3 \\
            & Strain       & 0.8   & 0.5  & 0    & 0    & 3 \\
        \bottomrule
    \end{tabular}
    \begin{tablenotes}
    \footnotesize
        \item \textit{Note.} Each attribute was rated across all available recordings (N=296)
    \end{tablenotes}
    \label{tab:capev_grbas_stats}
\end{table}

\begin{table}[!t]
\caption{Demographic Information of Speakers in Voice Samples.}
\begin{center}
\begin{tabular}{ccccc}
\toprule
\textbf{}&\multicolumn{2}{c}{Female/Male} & \multicolumn{2}{c}{Age (years)}  \\
\cline{2-5} 
\textbf{} & Samples& Percentage (\%)& Mean $\pm$ &Range \\
\midrule
Training& \text{143/83}& \text{63.3/36.7} & 46.31 $\pm$ 22.04 & 14-93 \\
Testing& \text{42/15}& \text{73.7/26.3} & 47.56 $\pm$ 21.04 & 18-90 \\
\bottomrule
\end{tabular}
\label{tab_demographic}
\end{center}
\end{table}

\begin{table}[!t]
\caption{Distribution of Training and Testing Samples for Utterance- and Patient-Level Assessments.}
\begin{center}
\begin{tabular}{c|cc|cc}
    \toprule
    \multirow{2}{*}{Evaluation Type} & \multicolumn{2}{c|}{PVQD-A} & \multicolumn{2}{c}{PVQD-S}  \\   
        & Training & Testing & Training & Testing \\
        \midrule
        Utterance-Level Evaluation & 226 & 57 & 1352 & 339 \\
        Patient-Level Evaluation & 226 & 57 & 226 & 57\\
        \bottomrule
\end{tabular}
\label{tab_prediction_levels}
\end{center}
\end{table}

\subsection{Data Split for Training and Testing}
To prevent data leakage and ensure generalizability, the PVQD dataset was split at the patient level, meaning that each speaker was exclusively assigned to either the training set or the test set. This ensured that the model was evaluating pathological voice quality on unseen speakers rather than memorized speech patterns, thus validating its ability to generalize beyond the training set. 

This approach follows that used in \cite{LightWeight} to ensure consistency with previous studies. The PVQD dataset originally contained 296 recordings, but 13 corrupted files were excluded, leaving a total of 283 valid samples. Specifically, 226 samples were used for training and 57 samples for testing. Table \ref{tab_demographic} provides the demographic distribution of speakers in the training and test sets, including gender ratio and age range. For a more comprehensive evaluation, vowel segments and connected-speech segments were extracted from each recording. Therefore, the PVQD dataset was divided into two subsets: PVQD-A (vowel-only subset), which contains /a/ vowel segments; and PVQD-S (speech-based subset), which consists of continuous speech segments. In this way, vowel phonation (PVQD-A) and connected-speech (PVQD-S) were evaluated independently. The number of samples in each subset is shown in Table~\ref{tab_prediction_levels}. Since each recording contains one /a/ vowel and multiple continuous speech segments, there are more samples of continuous speech than /a/ vowel.

\subsection{Feature Extraction}

All signals were resampled to 16 kHz before feature extraction. Two types of features were extracted. First, SFM embeddings from a pre-trained model (WavLM or Whisper) were used to capture phonetic and prosodic information. These embeddings were computed for both PVQD-A and PVQD-S to evaluate their effectiveness across different speech units. Second, LLDs, namely Jitter, Shimmer, and HNR were extracted using the Praat toolkit \cite{Praat} through the Parselmouth interface \cite{Parselmouth},\cite{Parselmouth2}. 

For the sustained-vowel subset (PVQD-A), recordings were made according to standard clinical instructions and were long enough to meet the requirement of $\ge$100 consecutive cycles for reliable perturbation analysis \cite{Patel2018}. Perturbation measures were calculated based on stable voiced portions automatically identified by Parselmouth \cite{Parselmouth2}, excluding onset and offset frames.

Given the clinical limitations of perturbation analysis, it may not be applicable to running speech due to frequent voiced/unvoiced transitions and turbulent noise. For the connected-speech subset (PVQD-S) consisting of six standardized CAPE-V sentences, perturbation analysis was applied only to voiced frames to obtain auxiliary acoustic descriptors for modeling, rather than as clinical perturbation measures. Notably, the primary acoustic representation used by VOQANet+ is the SFM embedding, which captures the major phonatory and spectral cues. LLDs provide only supplementary information.
Despite the limitations, combining these LLDs with SFM embeddings can provide complementary information and contribute to improved performance on sentence-based tasks.

\begin{table}[!t]
\caption{Noise Types and SNR Levels Used for Training, Seen-Test, and Unseen-Test.}
\begin{center}
\begin{tabular}{l|l|l}
    \toprule
     
         & Configuration & Details  \\
        \midrule
        Training & SNR & -5 dB, 0 dB, 5 dB, 10 dB  \\
        & Noise Type & White noise, Pink noise, \\
        & & Cafeteria babble, \& Cocktail party\\
        \hline
        Testing (Seen) & SNR & -5 dB, 0 dB, 5 dB, 10 dB  \\
        & Noise Type & White noise, Pink noise, \\
        & & Cafeteria babble, \& Cocktail party\\
        \hline
        Testing (Unseen) & SNR & 0 dB, 5 dB  \\
        & Noise Type & Brown noise \\
        & & Baby cry, \& Laughter\\
        \bottomrule
\end{tabular}
\label{tab_noise}
\end{center}
\end{table}

\subsection{Model Training}
All models were trained for 100 epochs using the AdamW optimizer (learning rate = 0.002, weight decay = 1e--5). Each model was trained and tested independently on the PVQD-A and PVQD-S subsets, and the performance on the two subsets is shown separately.




\subsection{Evaluation Criteria}
Model performance was evaluated using two metrics: Root Mean Squared Error (RMSE) and Pearson Correlation Coefficient (PCC). RMSE quantifies the average squared difference between the model output and the ground-truth perceptual score, with lower values indicating better performance. PCC measures the linear correlation between predicted and actual scores, with values closer to 1.0 indicating greater consistency with human ratings.

To reflect both fine-grained prediction accuracy and clinical relevance, we used both utterance-level and patient-level assessments. For patient-level scoring, we average the predictions from all utterances of a single speaker to arrive at a final perceptual rating. This approach mimics real-world clinical scenarios, where judgments are typically based on multiple utterances. This dual framework allows for a more comprehensive evaluation of the model's prediction performance, and the results are closely aligned with clinical practice.



\subsection{Noise Robustness Setup}
We introduced a noise-augmented version of the PVQD dataset to evaluate the robustness of the model under adverse acoustic conditions. Table~\ref{tab_noise} summarizes the noise types and signal-to-noise ratio (SNR) levels used for training, seen test, and unseen test scenarios.
Each set contains the original clean utterances. The training set was augmented with four noise types: white, pink (colored), cafeteria babble, and cocktail party (background), at SNRs of –5, 0, 5, and 10 dB, covering both stationary and non-stationary interference commonly used in speech robustness research. To evaluate generalization, unseen noise types were added during testing: brown noise (low-frequency), baby cry, and laughter, the latter two reflecting real-world non-speech vocalizations relevant to clinical and telehealth settings, where voice assessment may be performed in varied background settings. Fig.~\ref{waveformspectrogram} shows example waveforms and spectrograms for PVQD-A and PVQD-S under different noise conditions. The selected SNR range captures environments from challenging to moderately noisy, aligning with prior ASR and speech enhancement studies.

This experimental design ensures that the models are evaluated across a wide range of input conditions, including clean, noisy, vowel, and sentence-based speech, to comprehensively evaluate their generalizability, interpretability, and robustness.

\begin{table}[!t]
\caption{Performance Comparison of VOQANet and Baseline Models.\label{tab:res_pvqnet}}
\centering
\begin{tabular}{l|c|c|c|c|c}
\toprule
\multirow{2}{*}{Model} & \multirow{2}{*}{Feature} 
& \multicolumn{2}{c|}{PVQD-A} & \multicolumn{2}{c}{PVQD-S} \\

& & RMSE ↓ & PCC ↑  & RMSE ↓ & PCC ↑\\
\hline\hline
\multicolumn{6}{l}{CAPE-V Prediction}\\
\hline
Lin\cite{LightWeight} & HF & 15.22 &0.69&-&-\\
\hline
Lin \cite{LightWeight} & MFCC+MS & 14.76 & 0.64 & - & - \\
\hline
Lin \cite{LightWeight} & Waveform & 17.09 & 0.48 & - & - \\
\hline
Lin \cite{LightWeight} & W2V2 (Last) & 17.09 & 0.55 & - & - \\
\hline
Lin \cite{LightWeight} & HuBERT (Last) & 18.14 & 0.49 & - & - \\
\hline
Lin \cite{LightWeight} & WavLM (Last) & 20.23 & 0.33 & - & - \\
\hline
Lin \cite{LightWeight} & Whisper (Last) & 15.67 & 0.62 & - & - \\
\hline
VOQANet & HuBERT (Last) & 12.921 & 0.767 & 12.401 & 0.817 \\
\hline
VOQANet & Whisper (Last) & 10.514 & 0.803 & 10.546 & 0.843 \\
\hline
VOQANet & WavLM (Last) & \textbf{9.955}&\textbf{0.838}&\textbf{9.756} & \textbf{0.847}\\
\hline\hline
\multicolumn{6}{l}{GRBAS
Prediction}\\
\hline
VOQANet & HuBERT (Last) & 0.451 & 0.738 & 0.432 & 0.778 \\
\hline
VOQANet & Whisper (Last) & \textbf{0.380} & 0.793 & 0.352 & 0.819 \\
\hline
VOQANet & WavLM (Last) & \textbf{0.380}&\textbf{0.795}&\textbf{0.322} & \textbf{0.833}\\
\bottomrule
\end{tabular}
\end{table}

\begin{table}[!t]
\caption{Performance Comparison of VOQANet Models Using Different SFMs and Representations. \label {tab:res_embedd}}
\centering
\begin{tabular}{c|c|c|c|c|c}
\toprule
 \multirow{2}{*}{Feature} &\multirow{2}{*}{SFM} 
& \multicolumn{2}{c|}{PVQD-A} & \multicolumn{2}{c}{PVQD-S} \\

& & RMSE ↓ & PCC ↑  & RMSE ↓ & PCC ↑\\
\hline\hline
\multicolumn{5}{l}{CAPE-V Prediction}\\
    \hline
     \multirow{2}{*}{Last} & HuBERT & 12.921 & 0.767 & 12.401 & 0.817\\
     
     & Whisper & 10.514 & 0.803 & 10.546 & 0.843\\
    
    & WavLM &9.955&0.838&9.756 & 0.847\\
    \hline
    \multirow{2}{*}{WS} 
    & HuBERT & 11.851 & 0.823 & 11.389 & 0.842\\
    & Whisper & \textbf{9.770} & 0.854 & 9.933 & 0.863\\
    & WavLM &9.891&\textbf{0.865}&\textbf{9.209} & \textbf{0.870}\\
\hline\hline    
\multicolumn{5}{l}{GRBAS Prediction}\\
    \hline
    \multirow{2}{*}{Last} 
    & HuBERT & 0.451 & 0.738 & 0.432 & 0.778\\
    & Whisper & 0.380 & 0.793 & 0.352 & 0.819\\
    
    & WavLM &0.380&0.795&0.322 & 0.833\\
    \hline
    \multirow{2}{*}{WS} 
    & HuBERT & 0.391 & 0.767 & 0.396 & 0.799\\
    & Whisper & 0.370 & 0.800 & 0.354 & 0.822\\
    & WavLM &\textbf{0.369}&\textbf{0.809}&\textbf{0.318} & \textbf{0.845}\\

\bottomrule
\end{tabular}
\end{table}

\begin{table*}[t]
\caption{Performance Comparison of VOQANet and VOQANet+.}
\centering
    \begin{tabular}{l|c|cc|cc|cc|cc}
    \toprule
    \multirow{2}{*}{Method} & \multirow{2}{*}{Feature} 
    & \multicolumn{4}{c|}{Utterance-Level} 
    & \multicolumn{4}{c}{Patient-Level} \\
    \cline{3-10}
    
    
    & & \multicolumn{2}{c|}{PVQD-A} & \multicolumn{2}{c|}{PVQD-S} 
    & \multicolumn{2}{c|}{PVQD-A} & \multicolumn{2}{c}{PVQD-S} \\ 
    
    & & RMSE ↓ & PCC ↑  & RMSE ↓ & PCC ↑ & RMSE ↓ & PCC ↑ & RMSE ↓ & PCC ↑ \\ 
    \hline\hline
    
    \multicolumn{10}{l}{CAPE-V Prediction}\\
    \hline

    \multirow{2}{*}{VOQANet} & Whisper (WS) & 9.770 & 0.854 & 9.933 & 0.863 & 11.473 & 0.848 & 10.212 & 0.870\\
    \cline{2-10}
    & WavLM (WS) & 9.891 & 0.865 & 9.209 & 0.870 & 9.720 & 0.864 & 7.765 & 0.901\\
    \hline
    \multirow{2}{*}{VOQANet+} & Whisper (WS) + JSH & 9.304 & 0.868 & 9.922 & 0.866 & 9.790 & 0.862 & 9.350 & 0.875\\
    \cline{2-10}
    & \textbf{WavLM (WS) + JSH} & \textbf{8.594} & \textbf{0.877} & \textbf{8.720} & \textbf{0.883} & \textbf{9.042} & \textbf{0.878} & \textbf{7.356} & \textbf{0.908}\\

    \hline\hline
    \multicolumn{10}{l}{GRBAS Prediction}\\
    \hline

    \multirow{2}{*}{VOQANet} & Whisper (WS) & 0.370 & 0.800 & 0.354 & 0.822 & 0.342 & 0.813 & 0.324 & 0.855\\
    \cline{2-10}
    & WavLM (WS) & 0.369 & 0.809 & 0.318 & 0.845 & 0.337 & 0.826 & 0.297 & 0.867\\
    \hline
    \multirow{2}{*}{VOQANet+} & Whisper (WS) + JSH & 0.367 & 0.822 & 0.344 & 0.835 & 0.349 & 0.828 & 0.343 & 0.858\\
    \cline{2-10}
    & \textbf{WavLM (WS) + JSH} & \textbf{0.364} & \textbf{0.830} & \textbf{0.307} & \textbf{0.854} & \textbf{0.332} & \textbf{0.839} & \textbf{0.289} & \textbf{0.874}\\

    \bottomrule
    \end{tabular}
\label{tab_res_all_pvqnetplus}
\end{table*}

\begin{table*}[t]
\caption{Cross-Validation Performance of VOQANet and VOQANet+.}
\centering
    \begin{tabular}{l|cc|cc|cc|cc}
    \toprule
    \multirow{2}{*}{Model} 
    & \multicolumn{4}{c|}{Utterance-Level} 
    & \multicolumn{4}{c}{Patient-Level} \\
    \cline{2-9}
    
    
    & \multicolumn{2}{c|}{PVQD-A} & \multicolumn{2}{c|}{PVQD-S} 
    & \multicolumn{2}{c|}{PVQD-A} & \multicolumn{2}{c}{PVQD-S} \\ 
    
    & RMSE ↓ & PCC ↑  & RMSE ↓ & PCC ↑ & RMSE ↓ & PCC ↑ & RMSE ↓ & PCC ↑ \\ 
    \hline\hline
    
    \multicolumn{9}{l}{CAPE-V Prediction}\\
    \hline

    VOQANet 
        & 9.656\smallpm{0.648}
        & 0.847\smallpm{0.046}
        & 8.532\smallpm{0.488}
        & 0.897\smallpm{0.026}
        & 9.459 \smallpm{0.312}
        & 0.852 \smallpm{0.029}
        
        & 7.343 \smallpm{0.491}
        & 0.919 \smallpm{0.018}
    \\
    \hline
    VOQANet+ 
        & 9.376\smallpm{0.228}
        & 0.870\smallpm{0.049}
        & 8.084\smallpm{0.488}
        & 0.938\smallpm{0.026}
        & 9.002 \smallpm{0.520}
        & 0.881 \smallpm{0.017}
        & 6.266 \smallpm{0.843}
        & 0.958 \smallpm{0.018}
    \\
    \hline\hline
    \multicolumn{9}{l}{GRBAS Prediction}\\
    \hline
    VOQANet 
        & 0.425\smallpm{0.018}
        & 0.820\smallpm{0.022}
        & 0.334\smallpm{0.023}
        & 0.857\smallpm{0.035}
        & 0.419 \smallpm{0.031}
        & 0.832 \smallpm{0.074}
        
        & 0.299 \smallpm{0.022}
        & 0.889 \smallpm{0.028}
    \\
    VOQANet+ 
        & 0.398\smallpm{0.032}
        & 0.835\smallpm{0.036}
        & 0.324\smallpm{0.016}
        & 0.868\smallpm{0.026}
        & 0.370 \smallpm{0.011}
        & 0.849 \smallpm{0.058}
        & 0.289 \smallpm{0.015}
        & 0.900 \smallpm{0.018}
    \\
    
    \bottomrule
    \end{tabular}
\label{tab_res_cv}
\end{table*}

\begin{table*}[t]
\caption{Robustness Evaluation of VOQANet and VOQANet+ under Seen and Unseen Noisy Conditions.}
\centering
    \begin{tabular}{l|c|cc|cc|cc|cc}
    \toprule
    \multirow{2}{*}{Evaluation Type} & \multirow{2}{*}{Method} 
    & \multicolumn{4}{c|}{Seen} 
    & \multicolumn{4}{c}{Unseen} \\
    \cline{3-10}
    
    
    & & \multicolumn{2}{c|}{PVQD-A} & \multicolumn{2}{c|}{PVQD-S} 
    & \multicolumn{2}{c|}{PVQD-A} & \multicolumn{2}{c}{PVQD-S} \\ 
    
    & & RMSE ↓ & PCC ↑  & RMSE ↓ & PCC ↑ & RMSE ↓ & PCC ↑ & RMSE ↓ & PCC ↑ \\ 
    \hline\hline
    
    \multicolumn{10}{l}{CAPE-V Prediction}\\
    \hline

    \multirow{2}{*}{Utterance-Level} & VOQANet (WavLM (WS)) & 9.981 & 0.832 & 10.279 & 0.843 & 10.555 & 0.807 & 10.627 & 0.809\\
    \cline{2-10}
    & VOQANet+ (WavLM (WS) + JSH) & 9.453 & 0.844 & 9.318 & 0.852 & 10.393 & 0.809 & 10.579 & 0.811\\
    \hline
    \multirow{2}{*}{Patient-Level} & VOQANet (WavLM (WS))& 9.948 & 0.844 & 8.429 & 0.878 & 11.326 & 0.828 & 9.066 & 0.868\\
    \cline{2-10}
    & VOQANet+ (WavLM (WS) + JSH) & 9.381 & 0.852 & 8.265 & 0.888 & 10.096 & 0.832 & 8.625 & 0.881\\

    \hline\hline
    \multicolumn{10}{l}{GRBAS Prediction}\\
    \hline

    \multirow{2}{*}{Utterance-Level} & VOQANet (WavLM (WS)) & 0.365 & 0.778 & 0.375 & 0.836 & 0.381 & 0.774 & 0.402 & 0.807\\
    \cline{2-10}
    & VOQANet+ (WavLM (WS) + JSH & 0.362 & 0.785 & 0.320 & 0.841 & 0.373 & 0.779 & 0.326 & 0.836\\
    \hline
    \multirow{2}{*}{Patient-Level} & VOQANet (WavLM (WS)) & 0.365 & 0.817 & 0.299 & 0.858 & 0.452 & 0.802 & 0.336 & 0.832\\
    \cline{2-10}
    & VOQANet+ (WavLM (WS) + JSH) & 0.386 & 0.827 & 0.293 & 0.865 & 0.348 & 0.819 & 0.313 & 0.855\\

    \bottomrule
    \end{tabular}
\label{tab_robustness_test}
\end{table*}


\begin{table}[!t]
\caption{Ablation Study on Different LLD Types Within The VOQANet+ Framework. \label {tab:res_cpp}}
\centering
\begin{tabular}{c|c|c|c|c}
\toprule
 \multirow{2}{*}{Feature} 
& \multicolumn{2}{c|}{PVQD-A} & \multicolumn{2}{c}{PVQD-S} \\

& RMSE ↓ & PCC ↑  & RMSE ↓ & PCC ↑\\
\hline\hline
\multicolumn{4}{l}{CAPE-V Prediction}\\
    \hline
      CPP & 13.194 & 0.750 & 12.804 & 0.765 \\
      JSH & 11.555 & 0.787 & 14.864 & 0.708 \\
      WavLM(WS) + CPP & 9.506 & 0.868 & 8.258 & 0.897\\
      \textbf{WavLM(WS) + JSH} & \textbf{9.042} & \textbf{0.878} & \textbf{7.356} & \textbf{0.908}\\
\hline\hline
\multicolumn{4}{l}{GRBAS Prediction}\\    
    \hline
    CPP & 0.423 & 0.714 & 0.418 & 0.717 \\
    JSH & 0.398 & 0.772 & 0.518 & 0.685 \\
    WavLM(WS) + CPP & 0.370 & 0.828 & 0.296 & 0.869\\
    \textbf{WavLM(WS) + JSH} & \textbf{0.332} & \textbf{0.839} & \textbf{0.289} & \textbf{0.874}\\

\bottomrule
\end{tabular}
\end{table}

\begin{table}[!t]
\caption{Generalization Performance of VOQANet and VOQANet+ on the Saarbrücken Voice Database (SVD).\label{tab:res_svd}}
\centering
\begin{tabular}{c|c|c|c}
\toprule
Subset & Accuracy & F1-Score & AUC \\
\hline\hline
\multicolumn{4}{l}{VOQANet (WavLM (WS))}\\
\hline
SVD-A 
        & 0.7335\smallpm{0.0365} 
        & 0.7233\smallpm{0.0198}
        & 0.7432\smallpm{0.0459}\\ 
SVD-S  
        & 0.8375\smallpm{0.0264} 
        & 0.8064\smallpm{0.0335} 
        & 0.8455\smallpm{0.0262} \\ 
\hline\hline
\multicolumn{4}{l}{VOQANet+ (WavLM (WS) + JSH)}\\
\hline
SVD-A 
        & 0.8365\smallpm{0.0191} 
        & 0.8083\smallpm{0.0340} 
        & 0.8437\smallpm{0.0183} \\ 
SVD-S  
        & 0.8690\smallpm{0.0204} 
        & 0.8428\smallpm{0.0372}
        & 0.8794\smallpm{0.0243} \\ 
\bottomrule
\end{tabular}
\end{table}

\section{Results and Discussion}
\subsection{Comparison of VOQANet with the Baseline}
Table~\ref{tab:res_pvqnet} presents the utterance-level performance of VOQANet using different SFM embeddings (last-layer HuBERT, WavLM, and Whisper) for CAPE-V and GRBAS prediction. Both HuBERT and WavLM are trained through SSL, whereas Whisper follows a semi-supervised training strategy. We compare VOQANet with the methods of Lin et al.~\cite{LightWeight}, which evaluate traditional and neural regressors using LLDs, MFCC+MS, raw waveform features, and embeddings from pre-trained models such as W2V2, HuBERT, WavLM, and Whisper. Their evaluation focuses on CAPE-V prediction on the PVQD-A. Their results show that traditional features (HF and MFCC+MS) outperform most SFM features, with Whisper (Last) being the best SFM baseline (RMSE 14.76, PCC 0.69). However, their best results are significantly worse than those of VOQANet, suggesting that these traditional features are not sufficient to model the complex acoustic patterns relevant to pathological voice assessment. Notably, Lin et al. used noise-augmented training, whereas VOQANet in this experiment was trained only on clean PVQD-A data, yet still achieved far stronger results: RMSE to 12.921 (HuBERT (Last)), 10.514 (Whisper (Last)), and 9.955 (WavLM (Last)), with WavLM achieving the lowest RMSE and highest PCC (0.838).

On PVQD-S, VOQANet with WavLM (Last) also surpasses Whisper and HuBERT, suggesting that WavLM is particularly well suited to capture prosodic and phonetic nuances in continuous speech. Similar improvements were observed for GRBAS prediction on both subsets. Although GRBAS improvements appear smaller due to its narrower rating scale (0–3), they remain clinically meaningful.

\subsection{Comparison of SFM Types and Representations}
In this study, HuBERT, WavLM, and Whisper are used as backbone SFMs because of their excellent performance in speech quality assessment and the increasing relevance of SSL models in non-intrusive speech evaluation~\cite{chianghasa,zezario2023mosa}. To further analyze the effectiveness of SFM-based representations, we compare two adaptation strategies: last-layer features (Last) and weighted-sum aggregation (WS) for HuBERT, Whisper, and WavLM under utterance-level evaluation. As shown in Table \ref{tab:res_embedd}, WS representation consistently outperforms the last-layer representation in all configurations. For instance, in GRBAS prediction on PVQD-A, WavLM (WS) yields lower RMSE (0.369 vs. 0.380) and higher PCC (0.809 vs. 0.795). This aligns with prior work~\cite{WavLM,HuBERT} showing that aggregating multi-layer transformer features is more effective than relying solely on the final layer. The last layer primarily captures high-level linguistic and semantic representations, while the intermediate layers retain lower-level acoustic and phonatory information such as spectral smoothness, perturbation, and harmonic balance, which are essential for pathological voice analysis. The WS aggregation effectively integrates these multi-level features, enabling the model to leverage both global and fine-grained information for more accurate prediction of perceptual voice quality. Similar advantages have been reported in recent audio quality studies~\cite{Wisnu2025}. This finding supports the motivation for adopting WS aggregation in subsequent experiments using WavLM embeddings.

Table~\ref{tab:res_embedd} also shows that WavLM outperforms HuBERT and Whisper on both PVQD-A and PVQD-S. For CAPE-V prediction on PVQD-S, WavLM (WS) achieves lower RMSE (9.209 vs. 9.933) and higher PCC (0.870 vs. 0.863) than Whisper, with even larger gains for GRBAS (PCC 0.845 vs. 0.822). These findings confirm that WavLM's denoising pretraining and fine-grained acoustic modeling have advantages in disordered voice settings, especially when handling longer or more variable connected-speech contexts. Consequently, WavLM (WS) is adopted for subsequent experiments, which provides the most informative and powerful representation. Although prior studies identified HuBERT as particularly reliable for distinguishing normal from pathological voices~\cite{koudounas2024}, our findings show that Whisper and especially WavLM also deliver strong performance. This supports the clinical value of developing multiple reliable AI systems, as different SFMs may capture complementary acoustic cues relevant to dysphonia assessment. Therefore, in the following discussions, we only report WavLM as the representative for SSL model.

\begin{figure}[!t]
  \centering

  \subfloat[PVQD-A — Utterance-Level]{
    \includegraphics[width=0.45\linewidth]{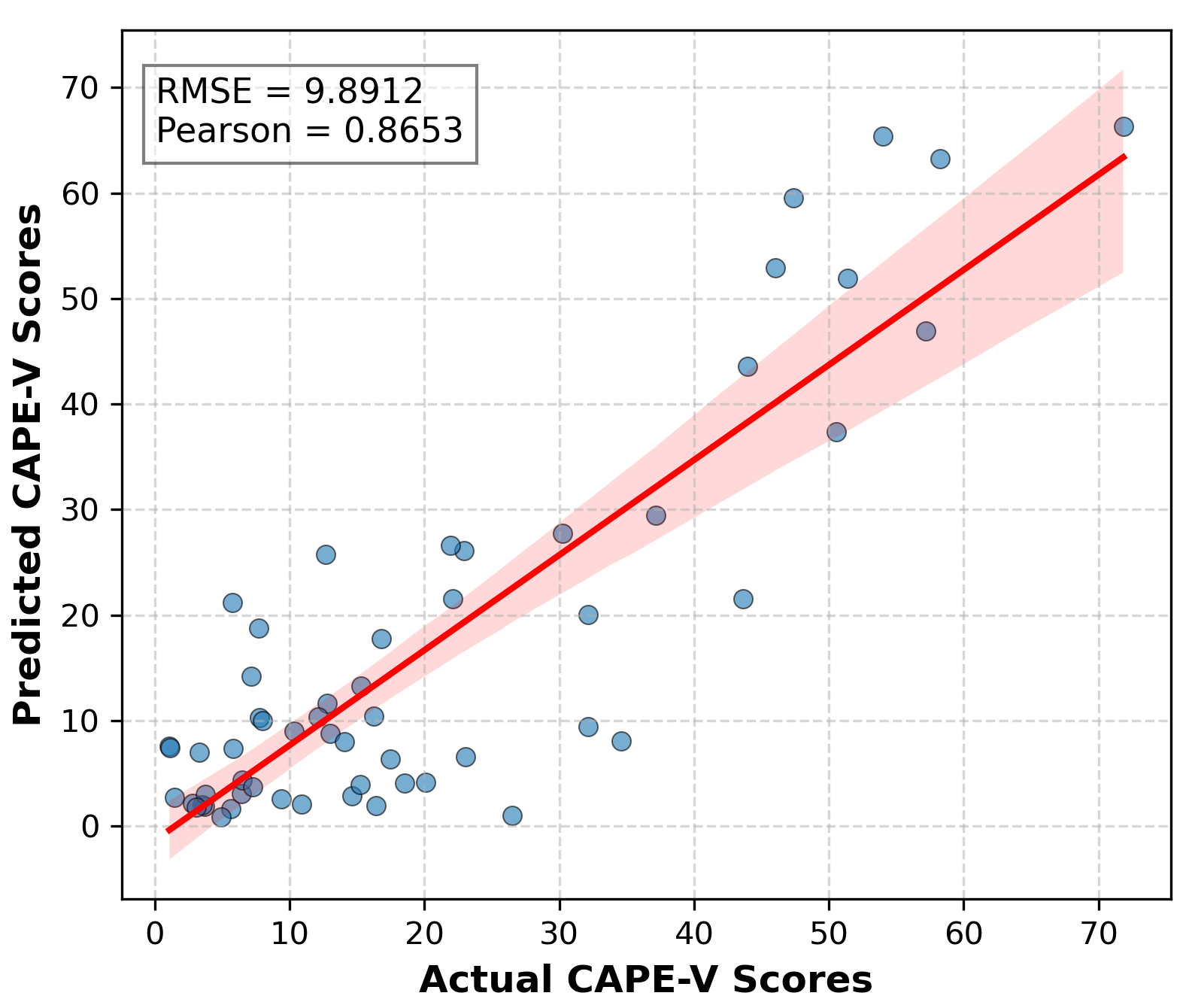}
  }\hspace{0.01\linewidth}  
  \subfloat[PVQD-S — Utterance-Level]{
    \includegraphics[width=0.45\linewidth]{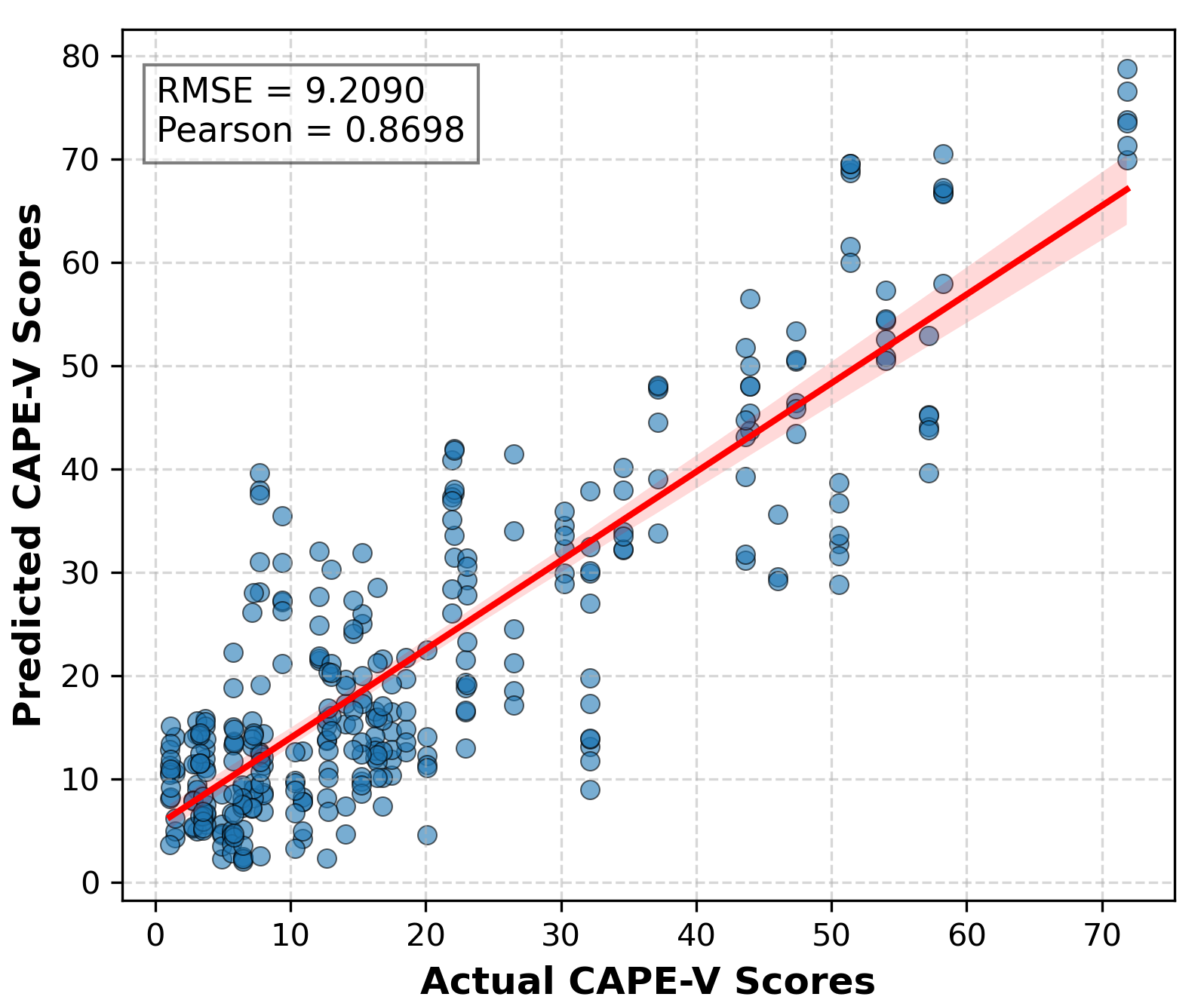}
  }

  \vspace{-0.5em} 

  \subfloat[PVQD-A — Patient-Level]{
    \includegraphics[width=0.45\linewidth]{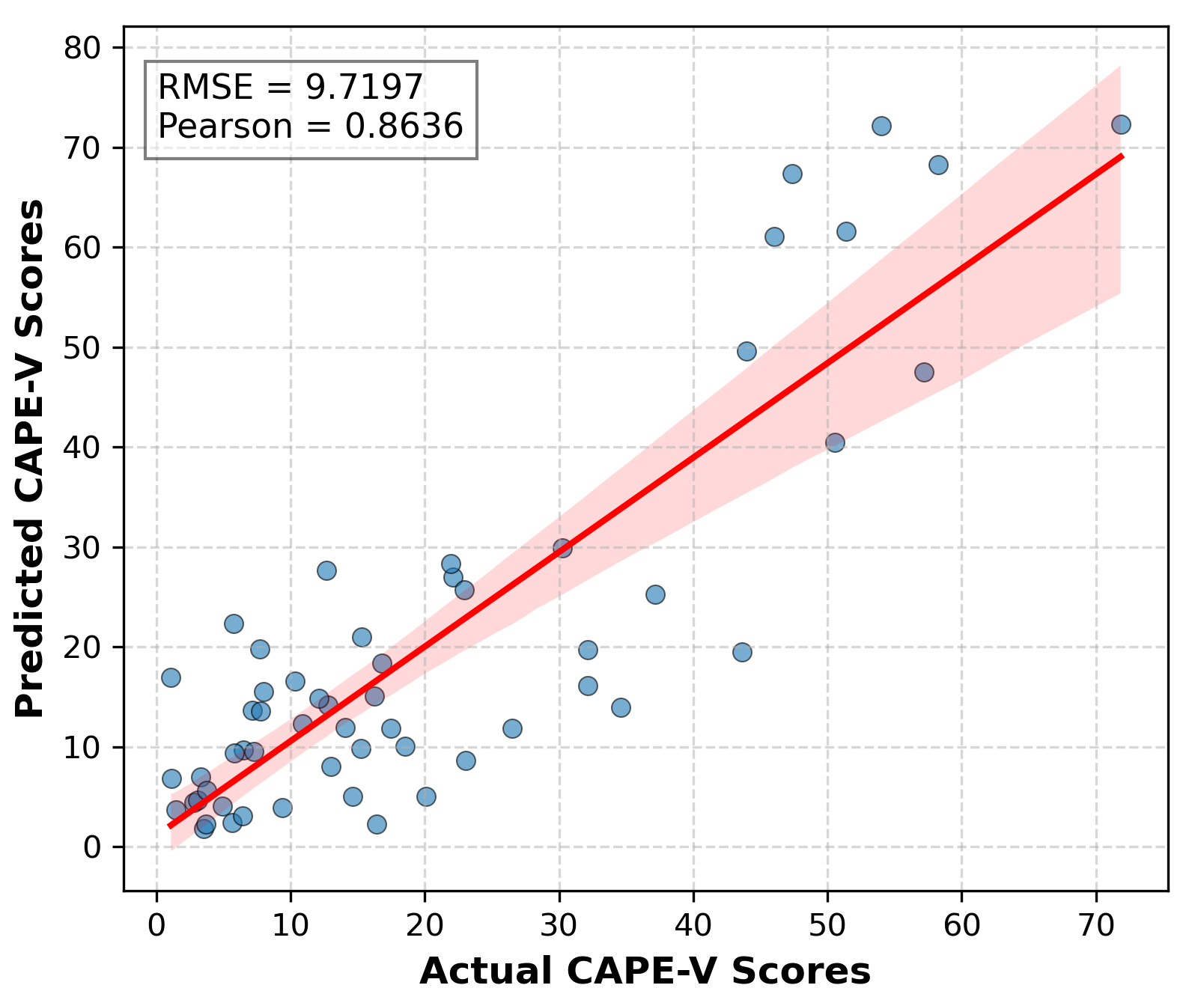}
  }\hspace{0.01\linewidth}
  \subfloat[PVQD-S — Patient-Level]{
    \includegraphics[width=0.45\linewidth]{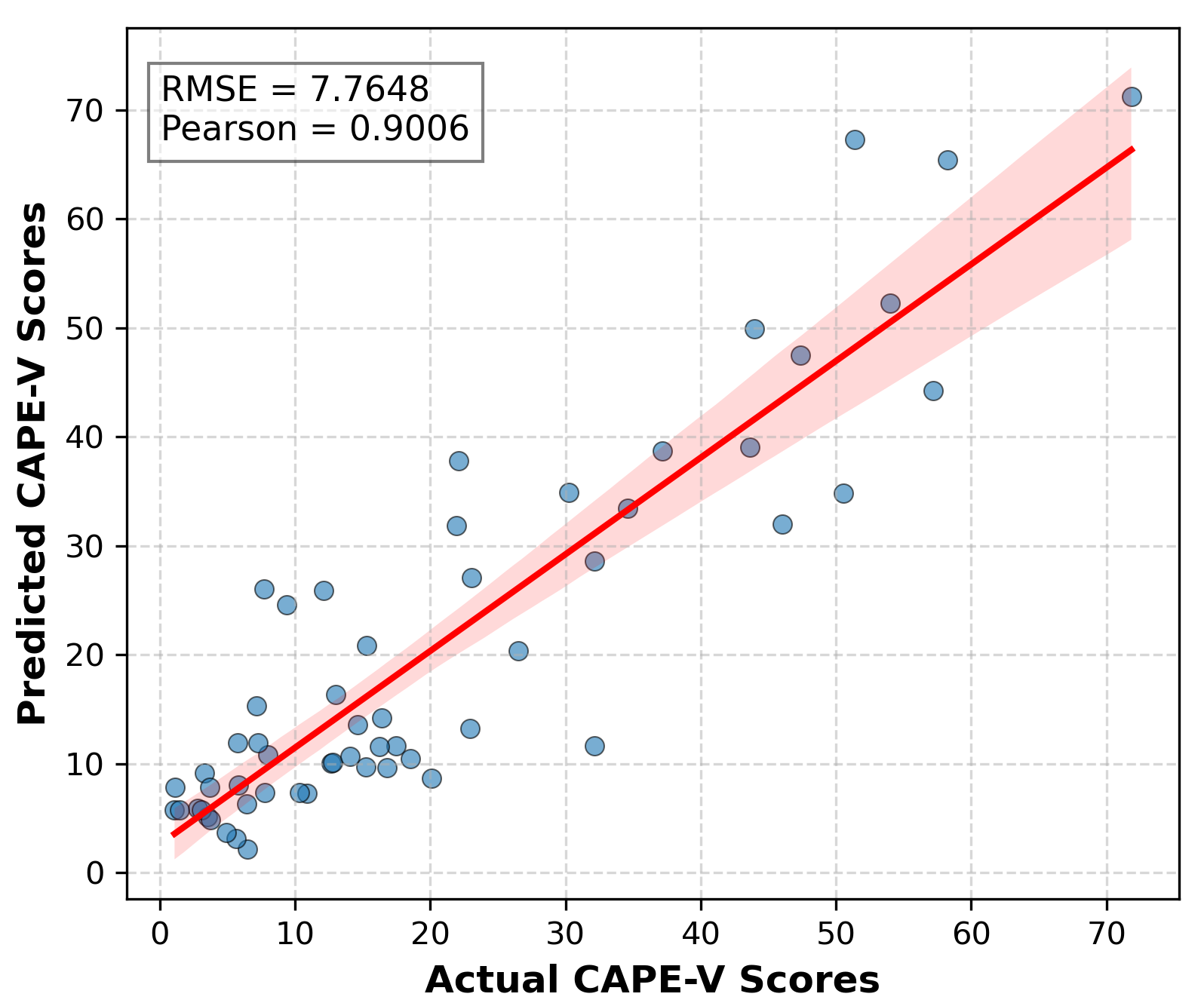}
  }

  \caption{Scatter plots of CAPE-V scores predicted by VOQANet with WavLM (WS) features versus actual scores. The top row shows utterance-level predictions on (a) PVQD-A and (b) PVQD-S, and the bottom row shows patient-level predictions on (c) PVQD-A and (d) PVQD-S.}
  \label{fig:pvqnet_capev_sctr}
\end{figure}

\begin{figure}[!t]
  \centering

  \subfloat[PVQD-A — Utterance-Level]{
    \includegraphics[width=0.45\linewidth]{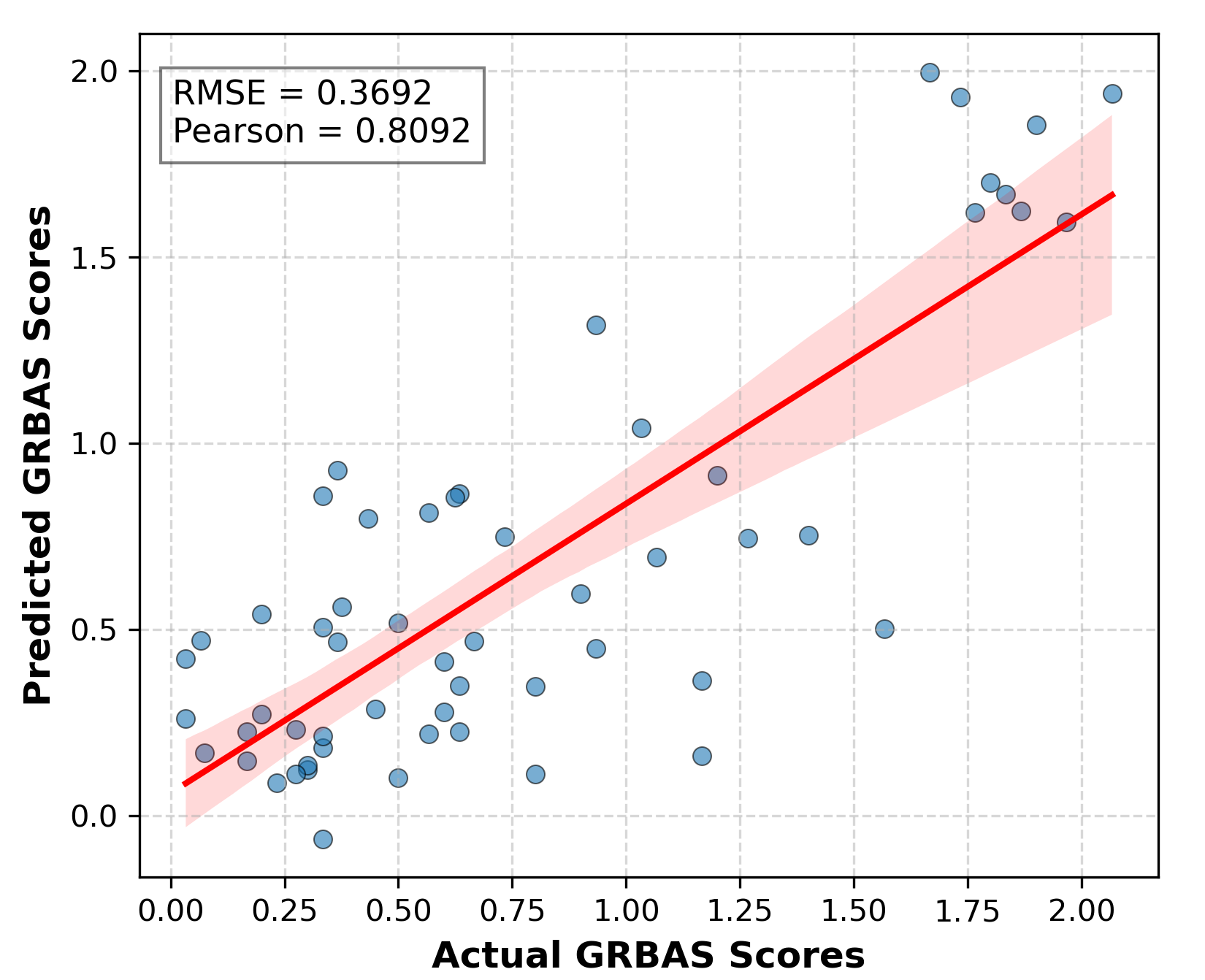}
  }\hspace{0.01\linewidth}  
  \subfloat[PVQD-S — Utterance-Level]{
    \includegraphics[width=0.45\linewidth]{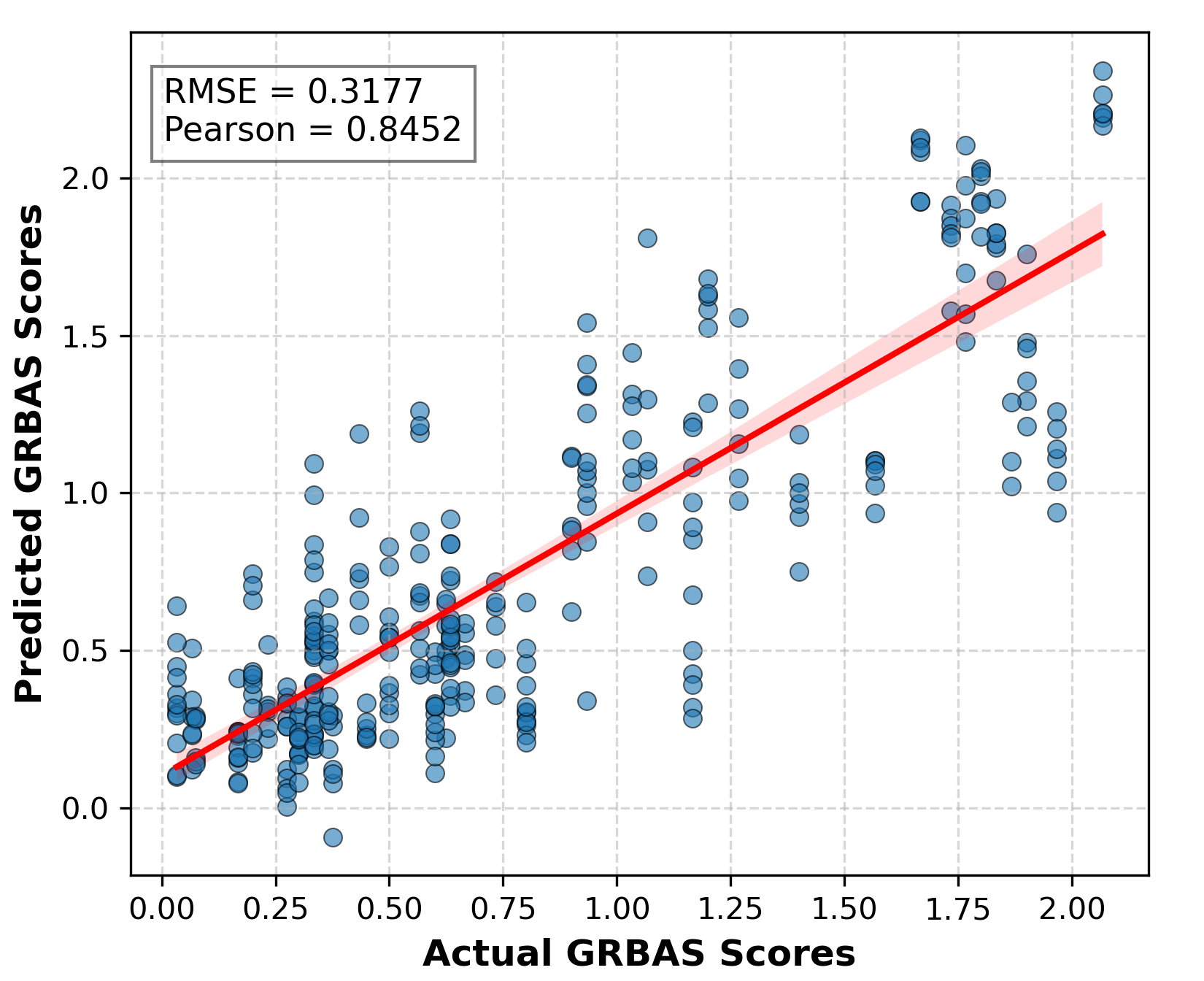}
  }

  \vspace{-0.5em} 

  \subfloat[PVQD-A — Patient-Level]{
    \includegraphics[width=0.45\linewidth]{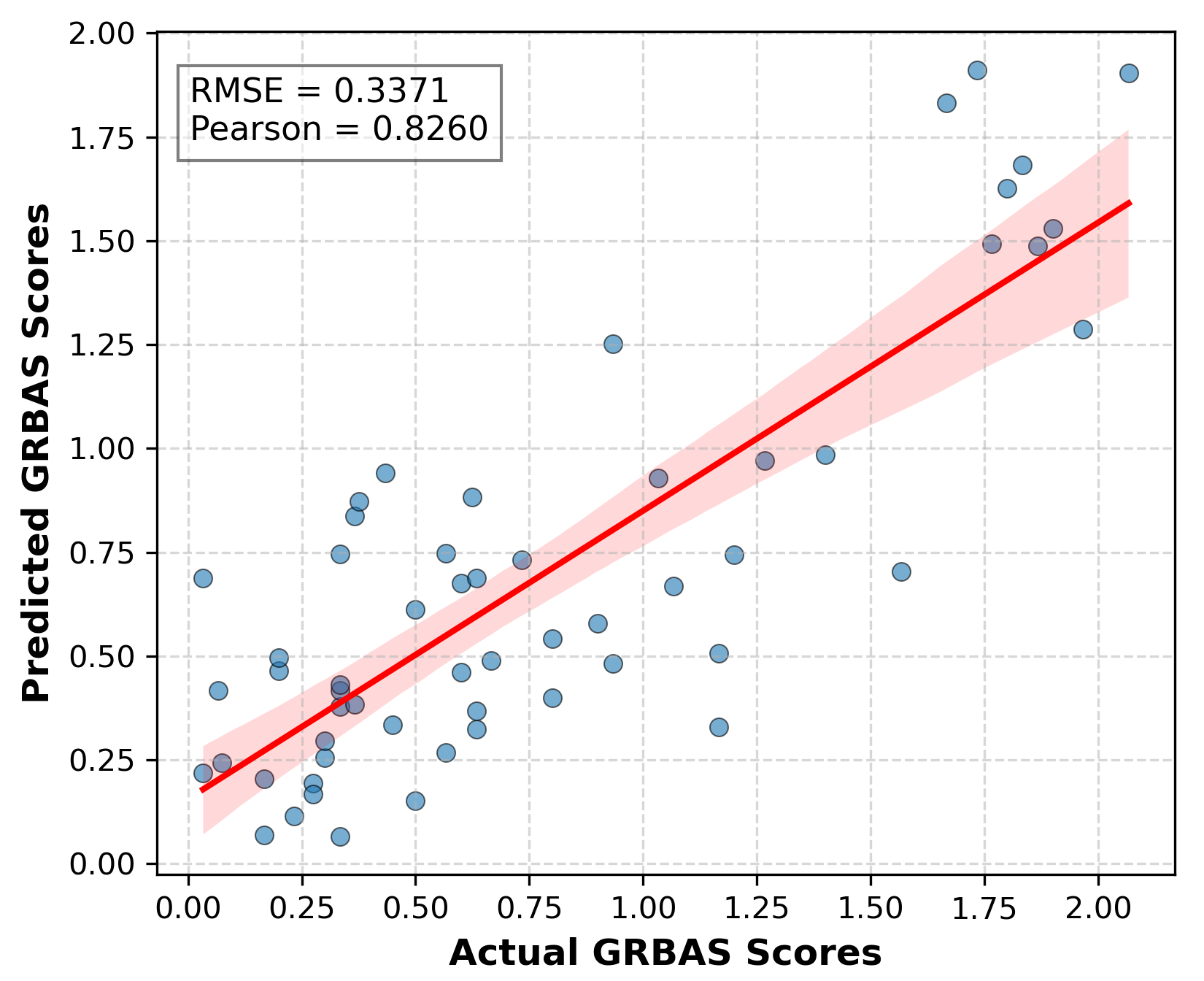}
  }\hspace{0.01\linewidth}
  \subfloat[PVQD-S — Patient-Level]{
    \includegraphics[width=0.45\linewidth]{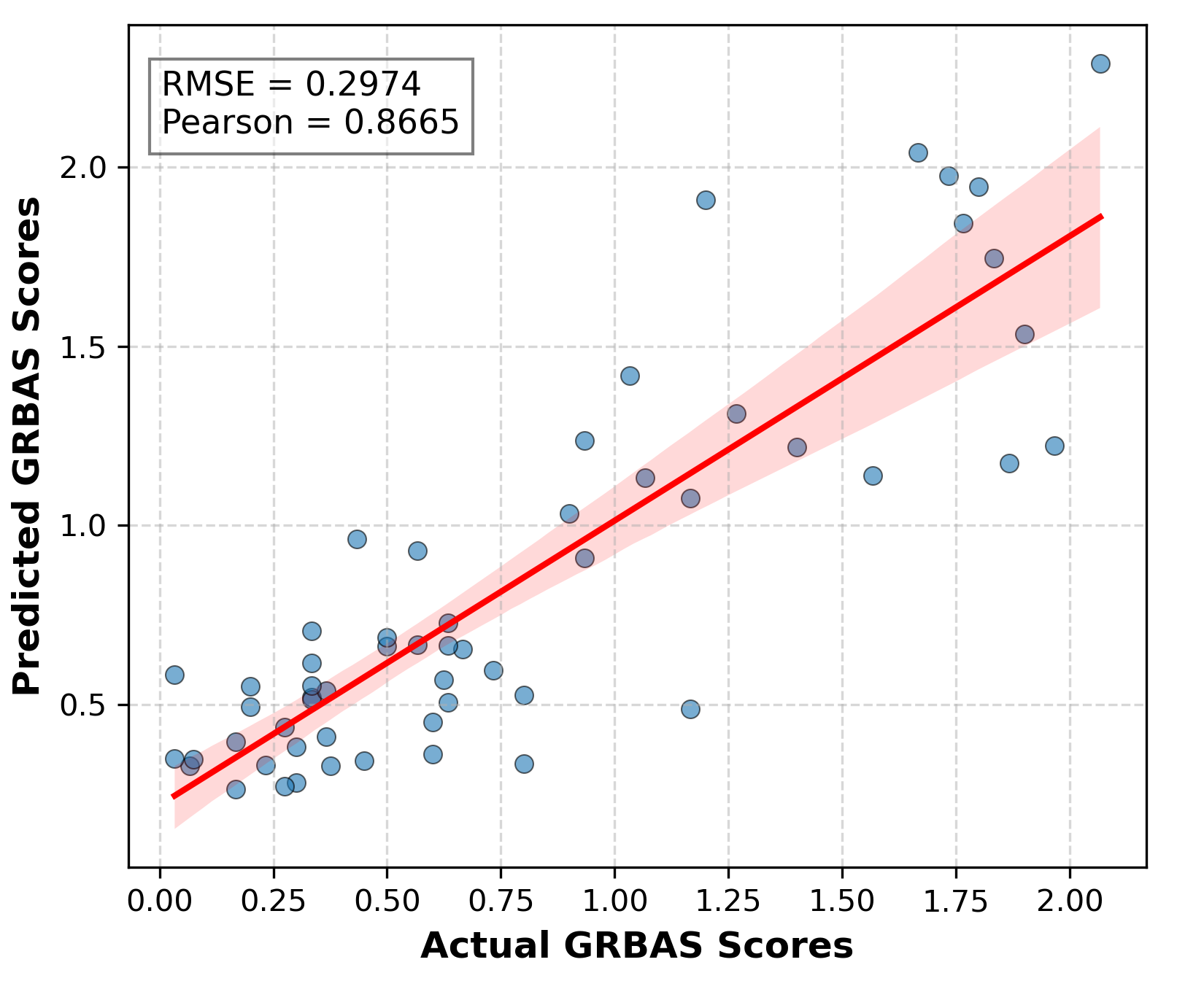}
  }

  \caption{Scatter plots of GRBAS scores predicted by VOQANet with WavLM (WS) features versus actual scores. The top row shows utterance-level predictions on (a) PVQD-A and (b) PVQD-S, and the bottom row shows patient-level predictions on (c) PVQD-A and (d) PVQD-S.}
  \label{fig:pvqnet_grbas_sctrr}
\end{figure}

\begin{figure}[!t]
  \centering

  \subfloat[PVQD-A — Utterance-Level]{
    \includegraphics[width=0.45\linewidth]{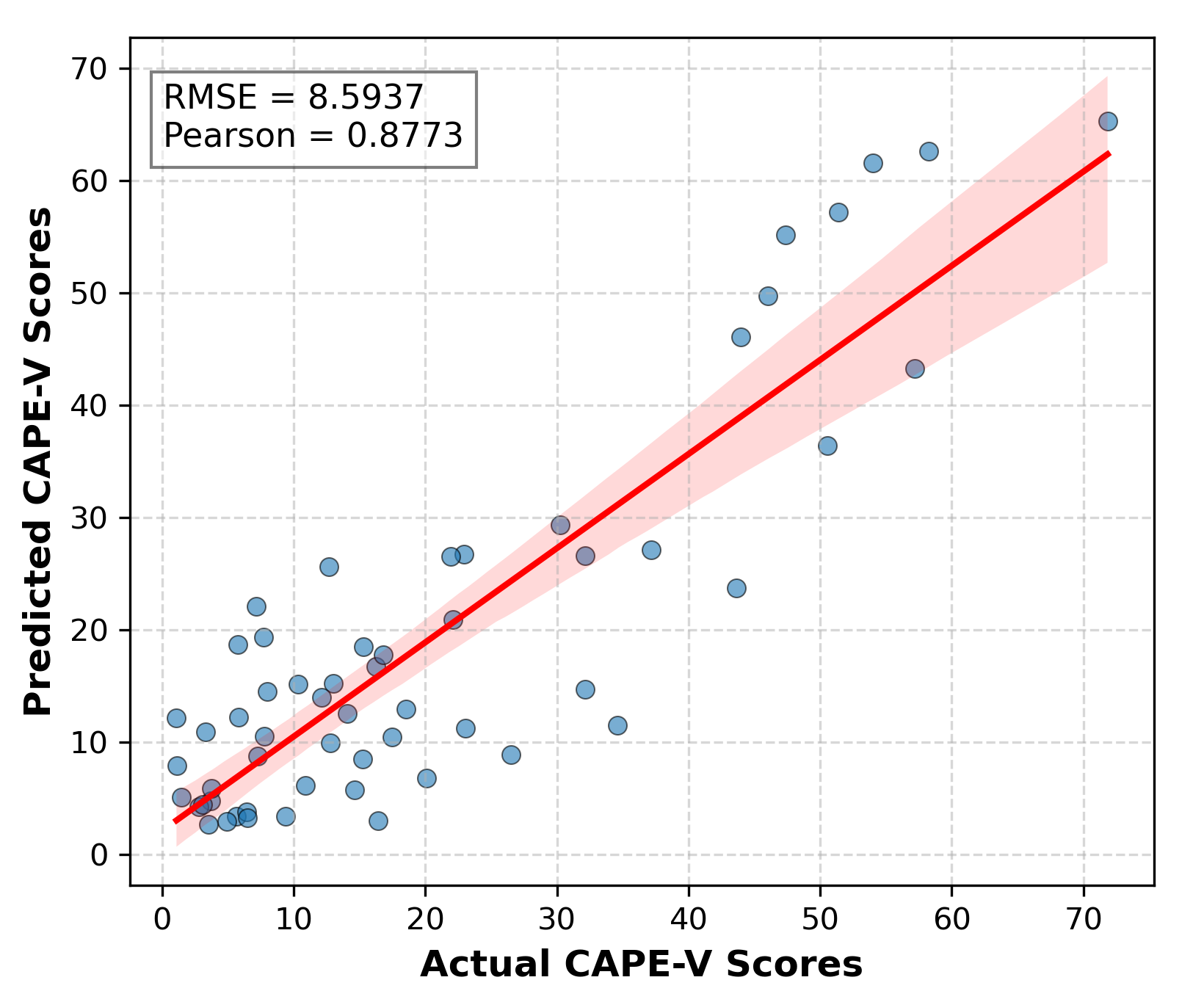}
  }\hspace{0.01\linewidth}  
  \subfloat[PVQD-S — Utterance-Level]{
    \includegraphics[width=0.45\linewidth]{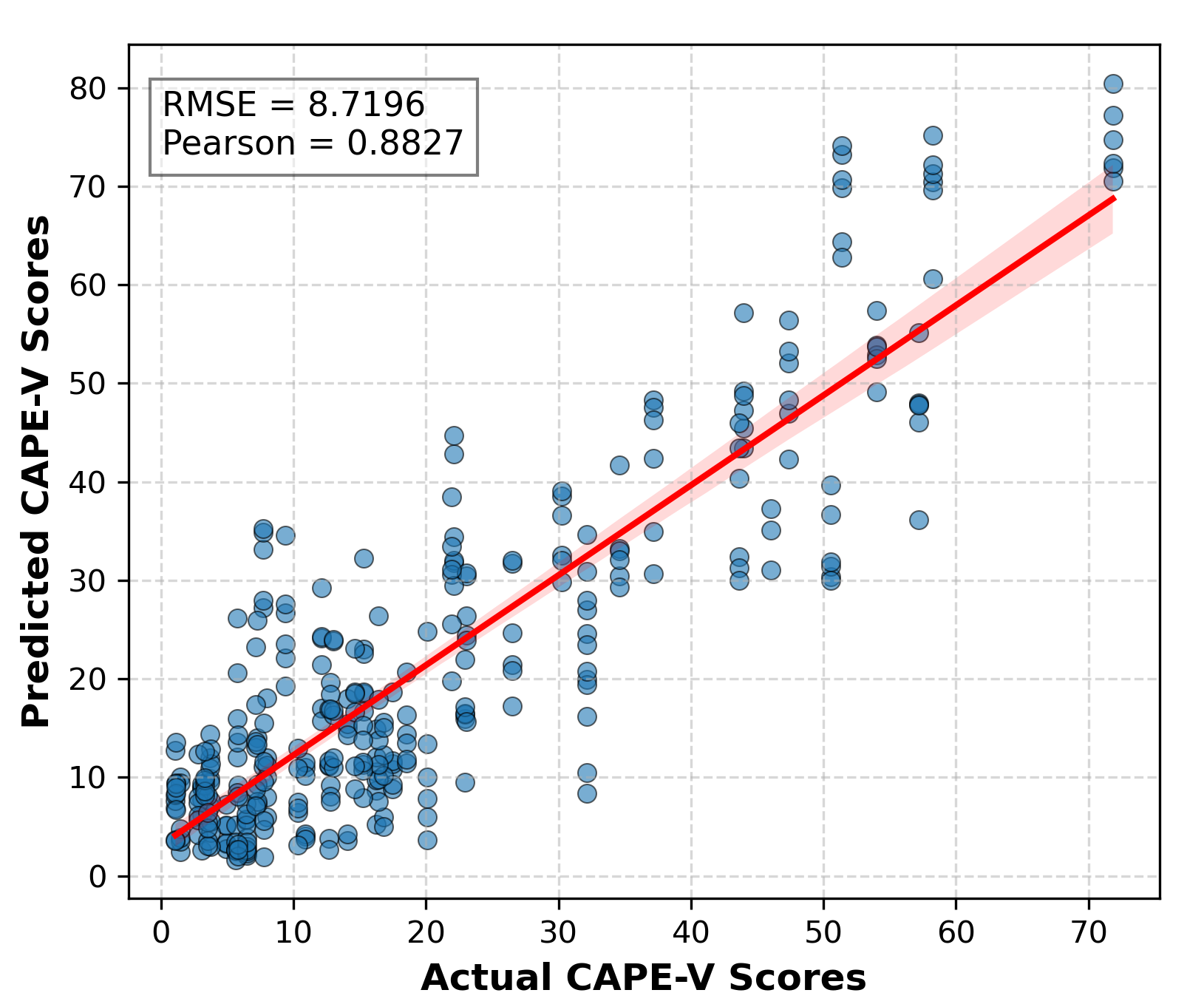}
  }

  \vspace{-0.5em} 

  \subfloat[PVQD-A — Patient-Level]{
    \includegraphics[width=0.45\linewidth]{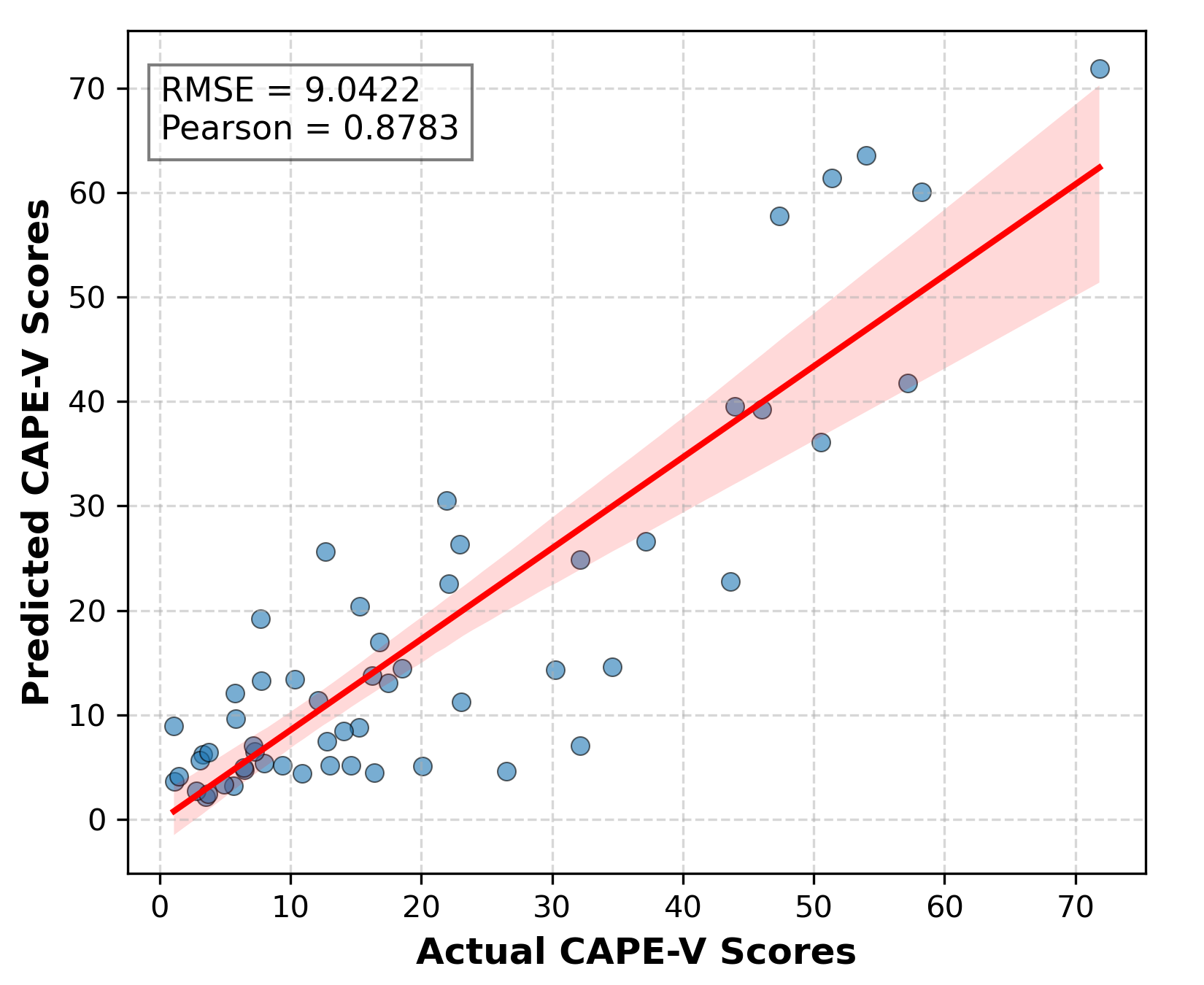}
  }\hspace{0.01\linewidth}
  \subfloat[PVQD-S — Patient-Level]{
    \includegraphics[width=0.45\linewidth]{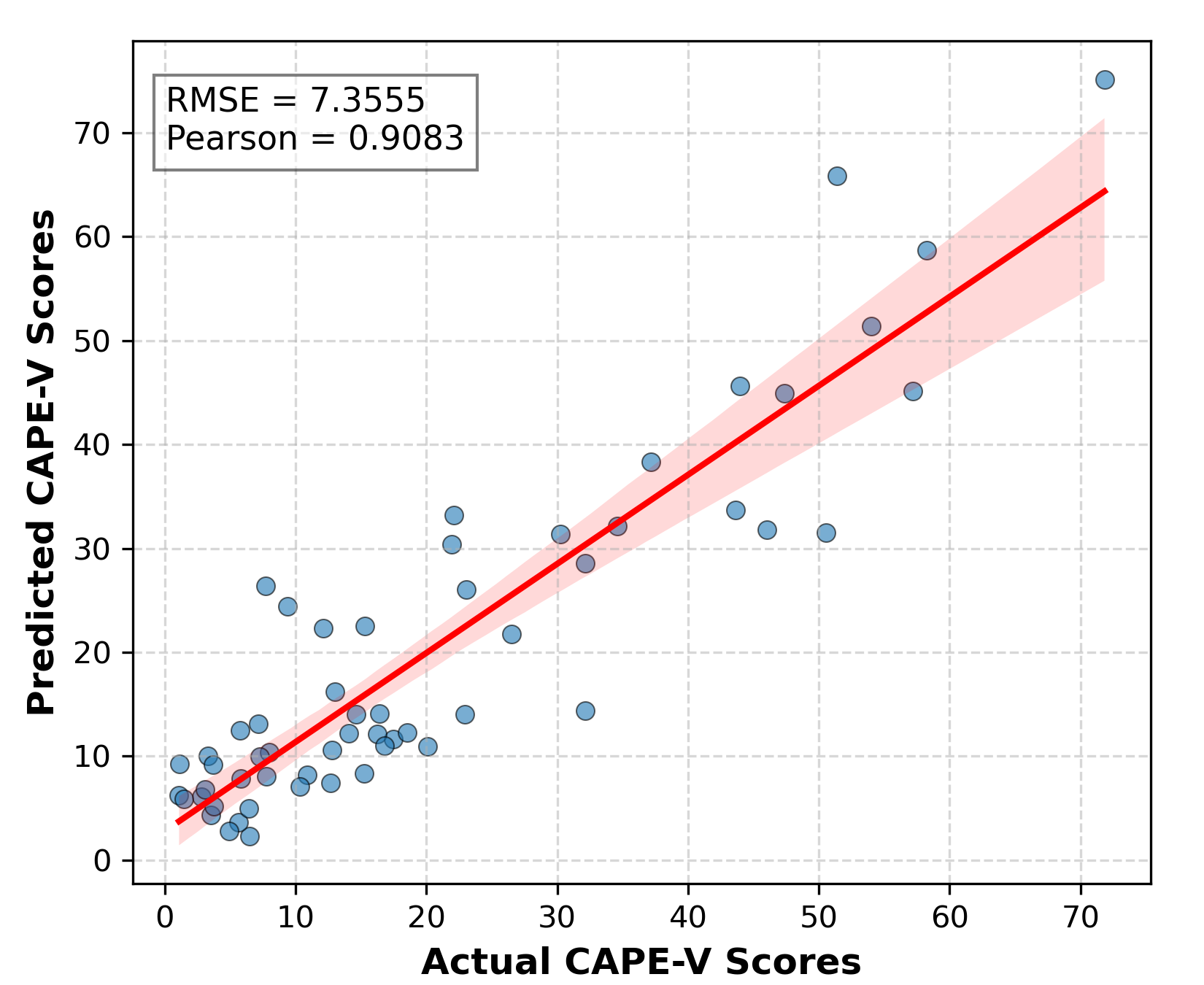}
  }

  \caption{Scatter plots of CAPE-V scores predicted by VOQANet+ with WavLM (WS) features and prosodic features (Jitter, Shimmer, and HNR). The top row shows utterance-level predictions on (a) PVQD-A and (b) PVQD-S, and the bottom row shows patient-level predictions on (c) PVQD-A and (d) PVQD-S.}
  \label{fig:sctr_pvqnetplus_capev}
\end{figure}

\begin{figure}[!t]
  \centering

  \subfloat[PVQD-A — Utterance-Level]{
    \includegraphics[width=0.45\linewidth]{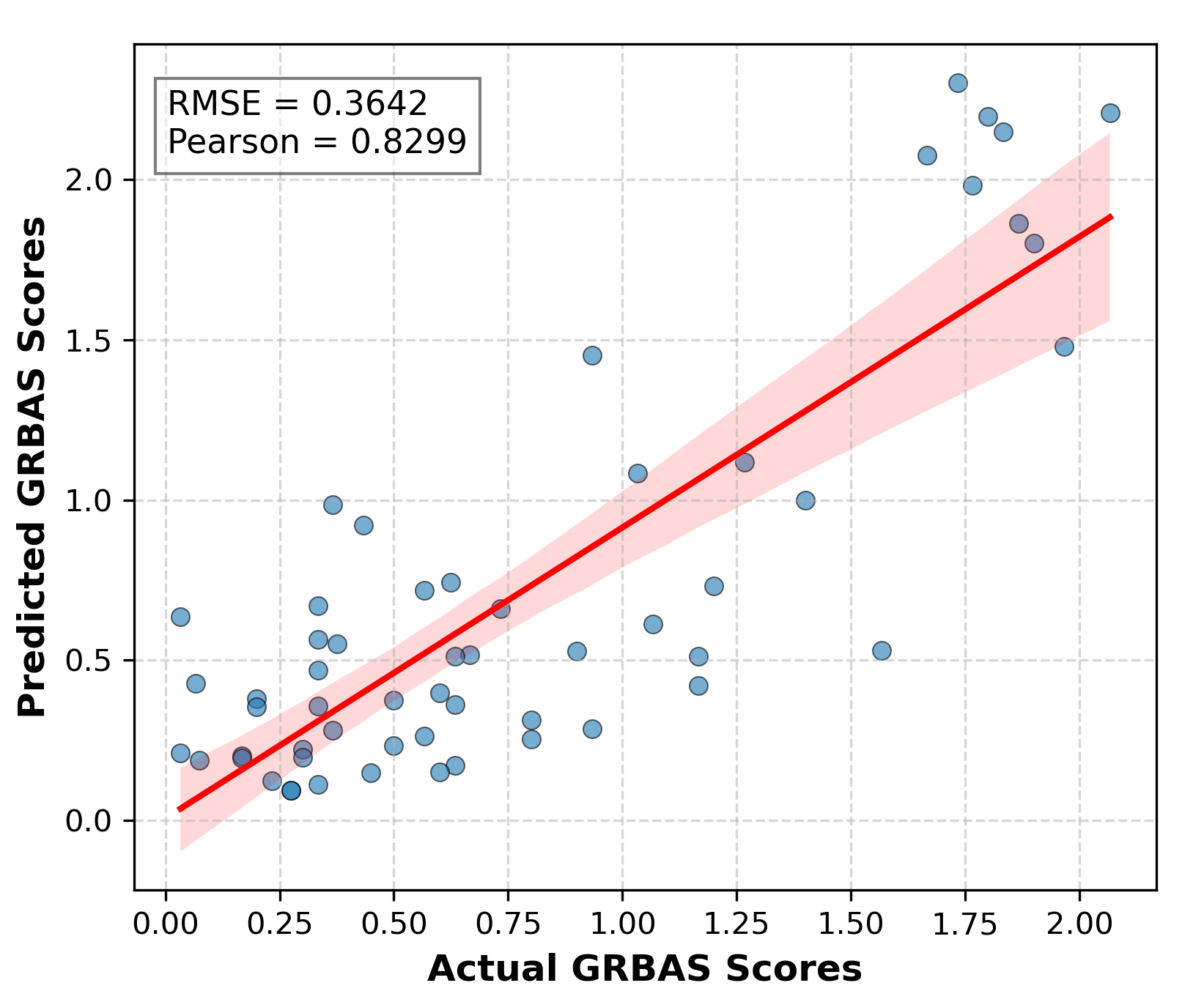}
  }\hspace{0.01\linewidth}  
  \subfloat[PVQD-S — Utterance-Level]{
    \includegraphics[width=0.45\linewidth]{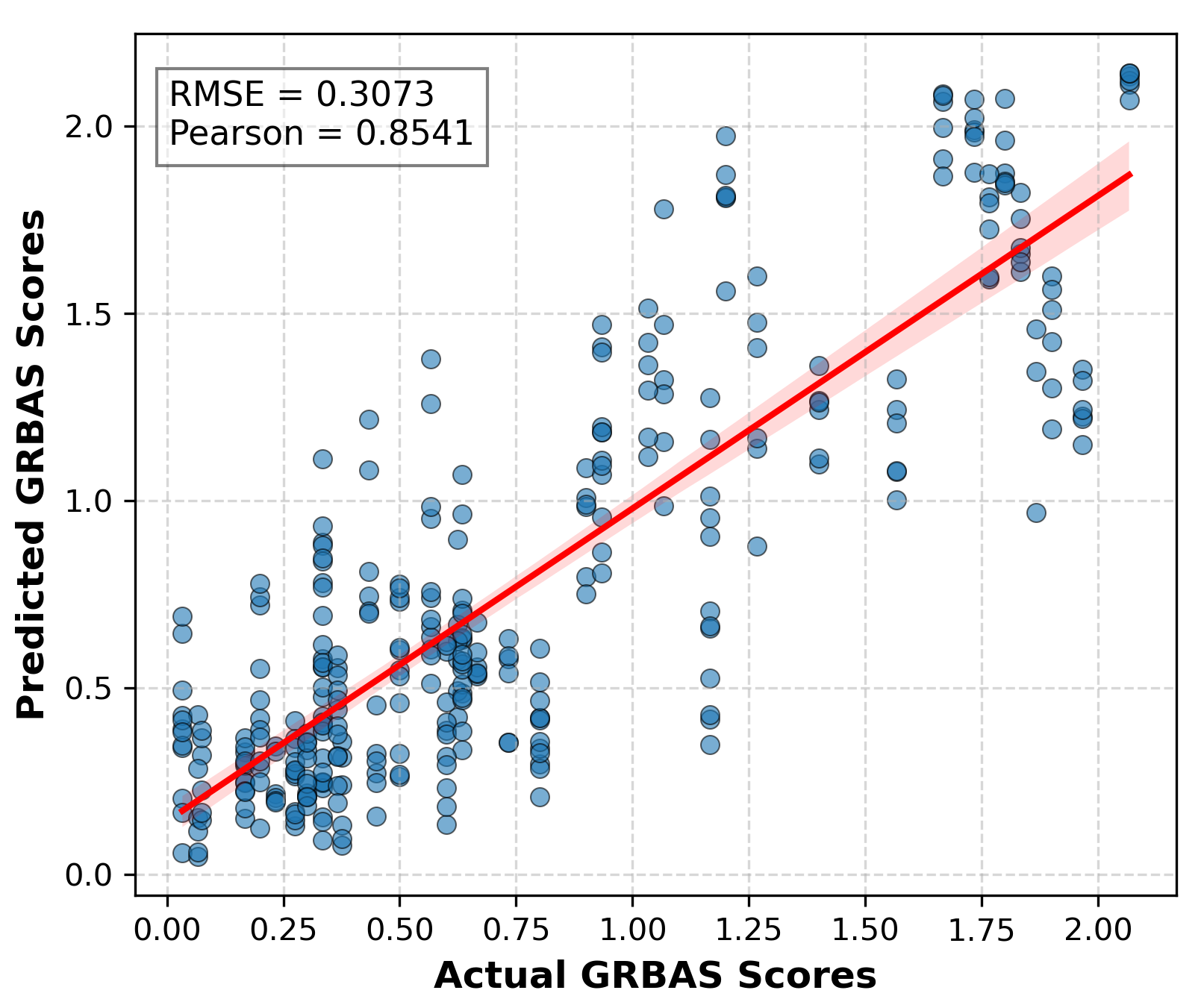}
  }

  \vspace{-0.5em} 

  \subfloat[PVQD-A — Patient-Level]{
    \includegraphics[width=0.45\linewidth]{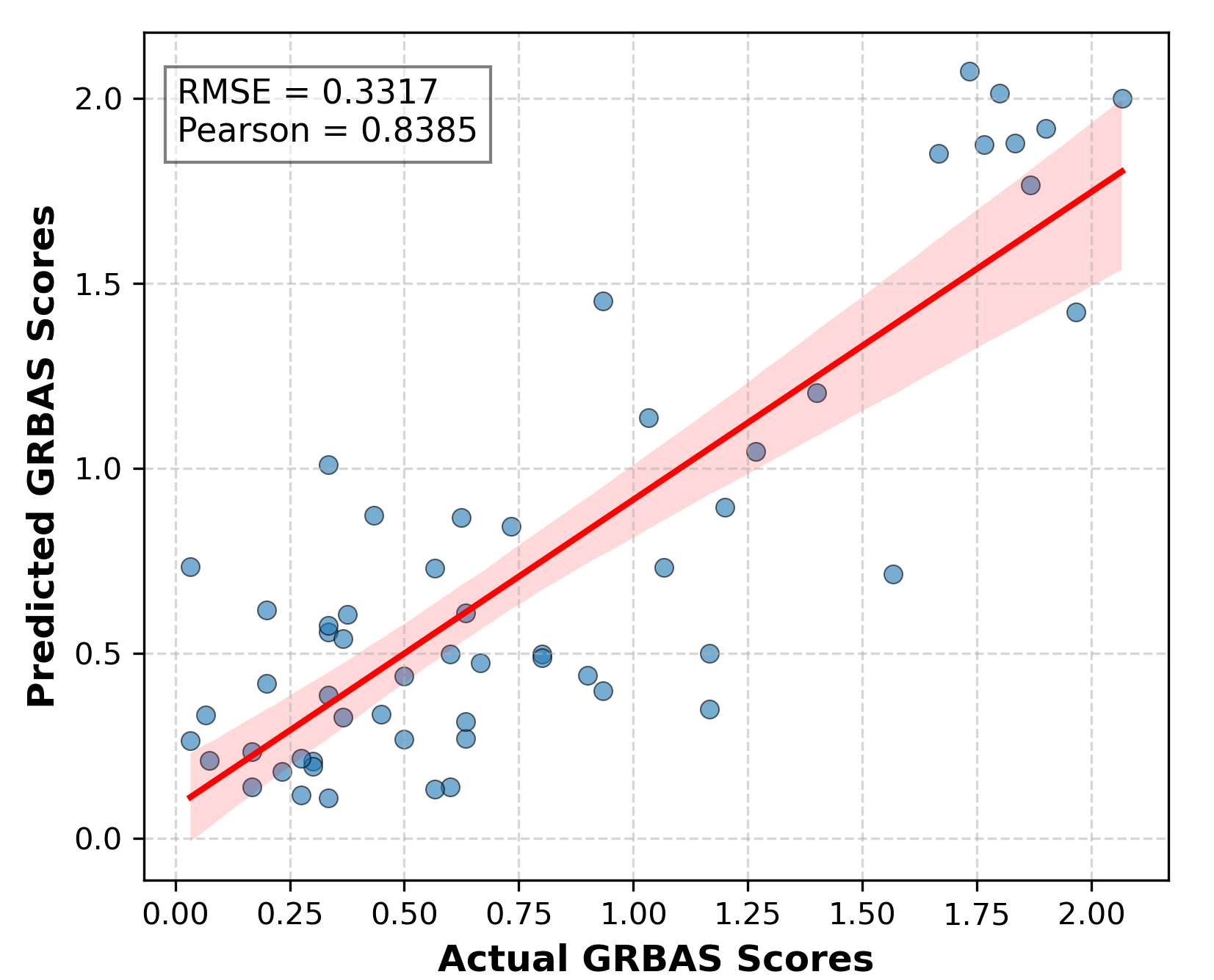}
  }\hspace{0.01\linewidth}
  \subfloat[PVQD-S — Patient-Level]{
    \includegraphics[width=0.45\linewidth]{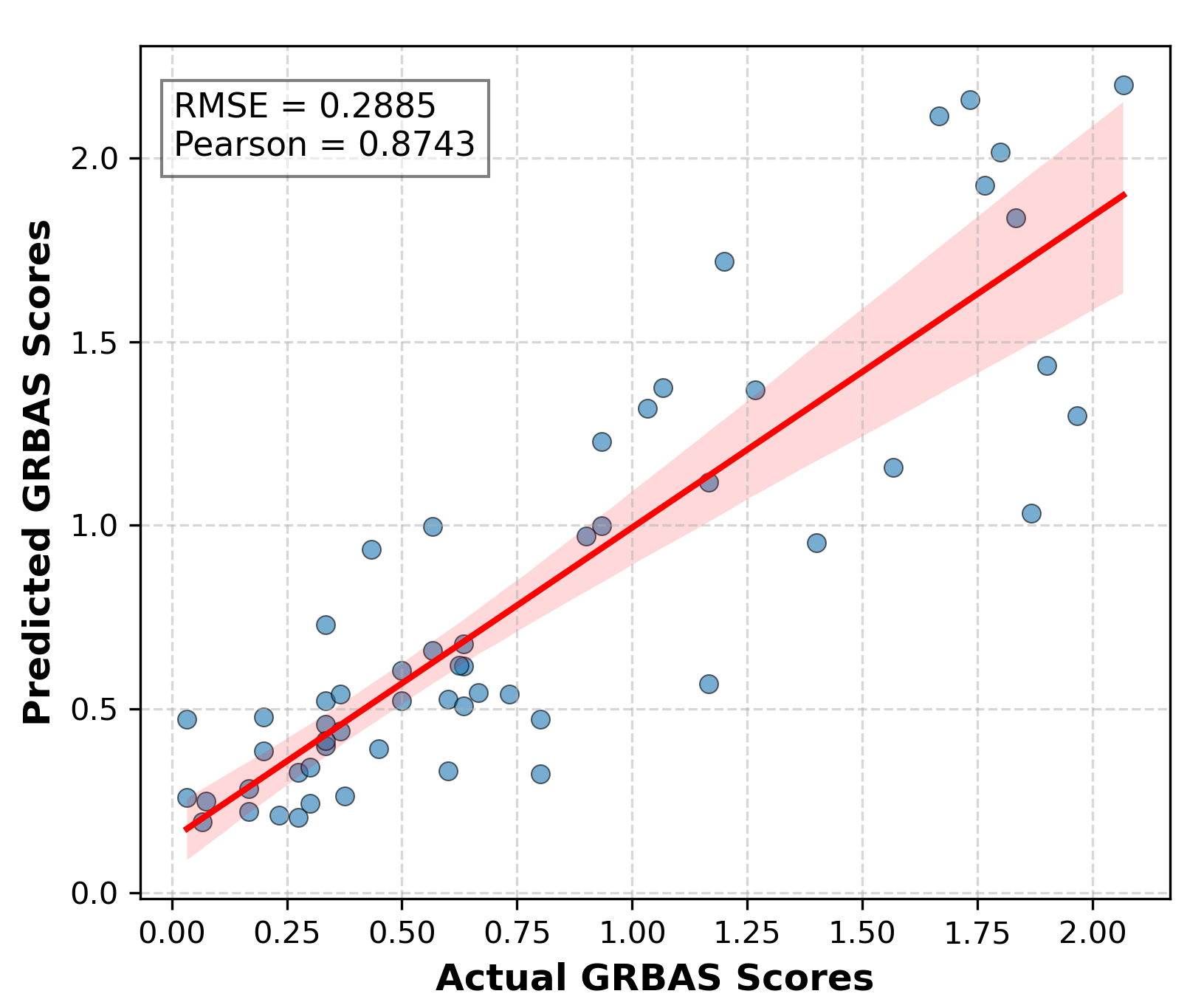}
  }

  \caption{Scatter plots of GRBAS scores predicted by VOQANet+ with WavLM (WS) features and prosodic features (Jitter, Shimmer, and HNR). The top row shows utterance-level predictions on (a) PVQD-A and (b) PVQD-S, and the bottom row shows patient-level predictions on (c) PVQD-A and (d) PVQD-S.}
  \label{fig:sctr_pvqnetplus_grbas}
\end{figure}

\begin{figure}[!t]
  \centering

  \subfloat[CAPE-V Prediction]{
    \includegraphics[width=0.45\linewidth]{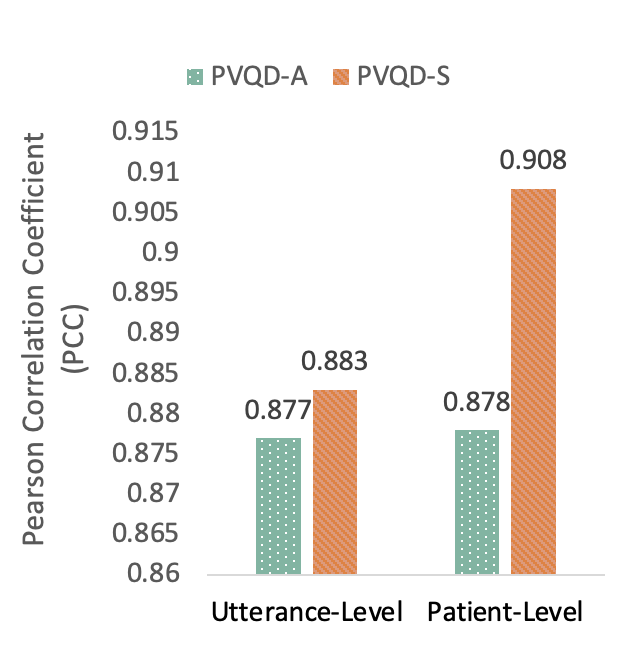}
  }\hspace{0.01\linewidth}  
  \subfloat[GRBAS Prediction]{
    \includegraphics[width=0.45\linewidth]{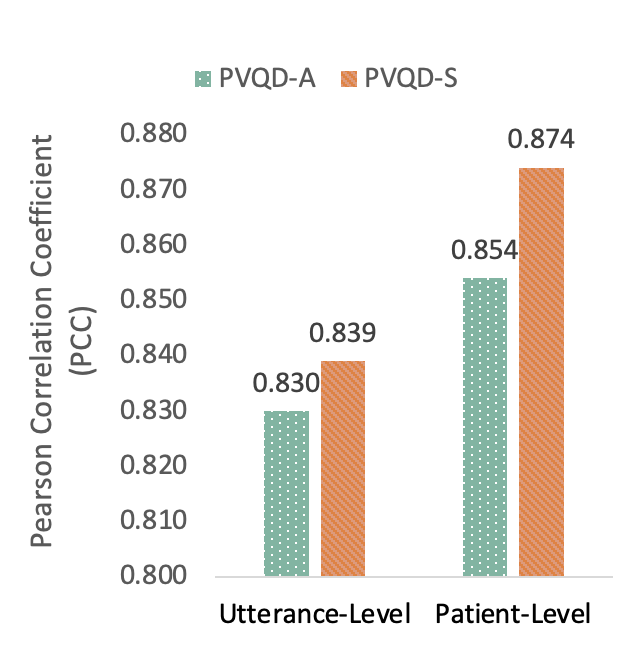}
  }

  \caption{Comparison of the performance of VOQANet+ (with WavLM (WS) + JSH features) for vowel-based predictions on PVQD-A and sentence-based predictions on PVQD-S.}
  \label{fig:pcc_barplot}
\end{figure}

\subsection{Comparison between VOQANet and VOQANet+} 
While SFM (e.g., WavLM) embeddings are effective in capturing high-level acoustic and prosodic information, they are not explicitly optimized for clinically salient voice quality traits. To address this limitation, VOQANet+ incorporates LLDs (jitter, shimmer, and HNR, or JSH for short), which correlate with pathological voice quality dimensions and are widely adopted in clinical voice analysis and speech processing systems for reflecting phonatory stability and noise~\cite{keller}. When combined with WavLM embeddings, these LLDs provide interpretable low-level signal-based information that complements deep learning-based representations. As shown in Table~\ref{tab_res_all_pvqnetplus}, VOQANet+ consistently achieves the lowest RMSE and highest PCC across all tasks and evaluation levels, outperforming VOQANet with SFM alone. The performance improvement is particularly prominent at the patient level, where predictions are aggregated over utterances from the same speaker to mimic real-world assessment. For example, in GRBAS prediction on PVQD-S, VOQANet+ improves the patient-level PCC from 0.867 to 0.874 and reduces RMSE from 0.297 to 0.289, indicating that adding LLDs helps to supplement low-level acoustic cues, such as irregularities in frequency or amplitude, that may not be adequately captured by SFM embeddings alone. 

The improvement achieved by VOQANet+ can be attributed to the complementary nature of the two feature domains. 
While SFM (WavLM) embeddings capture global prosodic and contextual representations learned from large-scale pretraining, they may overlook subtle perturbation-based cues that are clinically relevant for dysphonia characterization. 
In contrast, LLDs such as JSH explicitly encode cycle-to-cycle variations and noise-to-harmonic ratios—features closely aligned with perceptual dimensions like roughness, breathiness, and strain. 
By combining these representations, VOQANet+ benefits from both the rich contextual modeling of SFMs and the fine-grained phonatory information of LLDs, leading to predictions that are not only more accurate but also more interpretable in a clinical context.

Although acoustic measures are objective, they remain sensitive to speaker performance and recording variability, further motivating their combination with robust SFM embeddings in VOQANet+ to achieve more robust and consistent predictions. Fig.~\ref{fig:pvqnet_capev_sctr} visualizes the CAPE-V scores predicted by VOQANet versus the actual scores, while Fig.~\ref{fig:pvqnet_grbas_sctrr} does so for GRBAS. Fig.~\ref{fig:sctr_pvqnetplus_capev} visualizes the CAPE-V scores predicted by VOQANet+ versus the actual scores, while Fig.~\ref{fig:sctr_pvqnetplus_grbas} does so for GRBAS. In each figure, the x-axis denotes the ground-truth scores assigned by expert raters, and the y-axis represents the model's predicted scores. The top row (e.g., Fig.~\ref{fig:pvqnet_capev_sctr}(a–b)) corresponds to utterance-level predictions, and the bottom row (e.g., Fig.~\ref{fig:pvqnet_capev_sctr}(c–d)) shows patient-level predictions obtained by averaging the prediction scores of different utterances of the same speaker. Each point represents an utterance (or speaker), the red line shows the regression fit, and the shaded area is the 95\% confidence interval. VOQANet+ shows tighter clustering near the diagonal, especially in the patient-level plots (e.g., Fig.~\ref{fig:pvqnet_capev_sctr}(d) vs. Fig.~\ref{fig:sctr_pvqnetplus_capev}(d) and Fig.~\ref{fig:pvqnet_grbas_sctrr}(d) vs. Fig.~\ref{fig:sctr_pvqnetplus_grbas}(d)), which indicates stronger prediction consistency and lower variance. These findings further demonstrate that LLDs can enhance the model’s ability to estimate perceptual ratings, especially in cases of higher severity, where expert judgments tend to be more variable. In summary, by combining interpretable LLDs with SFM embeddings, VOQANet+ improves the clinical relevance, trustworthiness, and robustness of pathological voice quality prediction.

\subsection{Utterance-Level Evaluation vs. Patient-Level Evaluation}
In clinical settings, auditory perceptual judgments are often made by listening to multiple utterances from a single speaker. To reflect this practice, we evaluate model performance at both the utterance level and the patient level by averaging the predictions for all utterances from the same speaker. As shown in Table~\ref{tab_res_all_pvqnetplus} and visualized in scatter plots (Figs.~\ref{fig:pvqnet_capev_sctr}, \ref{fig:pvqnet_grbas_sctrr}, \ref{fig:sctr_pvqnetplus_capev}, \ref{fig:sctr_pvqnetplus_grbas}), patient-level evaluation achieves higher PCC and lower RMSE than utterance-level evaluation. For instance, with VOQANet+ (with WavLM (WS) + JSH features) for CAPE-V prediction on PVQD-S, PCC improved from 0.883 at the utterance level to 0.908 at the patient level, and reduce RMSE from 8.720 to 7.356.

To further investigate the contribution of different speech types, we compare the results of VOQANet+ (with WavLM (WS) + JSH features) on the PVQD-A (vowel-based) and PVQD-S (sentence-based) subsets. As shown in Fig.~\ref{fig:pcc_barplot}, sentence-based predictions consistently outperform vowel-based predictions, especially at the patient level. This indicate that connected-speech provides richer prosodic and articulatory information, allowing the model to make more accurate and stable predictions. These results suggest that while sustained vowels are still clinically useful, sentence-level inputs can provide more contextual acoustic dynamics, such as stress, pitch variation, and connected phonation, which are highly informative for complex perceptual dimensions such as strain or roughness. It is important to note, however, that sustained vowels remain essential for analyzing the phonatory sound source with minimal articulatory interference, whereas connected-speech reflects additional suprasegmental and articulatory factors. Consequently, the two tasks represent complementary functional contexts in vocal analysis rather than interchangeable conditions.

Moreover, the advantage of sentence-based predictions is further amplified when LLDs are incorporated into the model (see the comparison of VOQANet and VOQANet+ in Table~\ref{tab_res_all_pvqnetplus}), suggesting a synergistic effect between rich SFM-derived representations and domain-informed acoustic features, especially in the capture of the characteristics of voice disorders. As shown in the next subsection, sentence-level input enables VOQANet+ to maintain stronger performance under seen and unseen noisy conditions. This resilience further highlights the clinical value of incorporating connected-speech into the automated voice assessment framework.

\subsection{Cross-Validation Performance of VOQANet and VOQANet+}

We performed a five-fold patient-disjoint cross-validation (CV) across all experiments at both the utterance and patient levels, ensuring speaker-independent data partitions. Each fold preserved the same ratio of training and validation samples as in the original dataset, thereby maintaining a consistent distribution of samples and perceptual ratings. This strategy provides a more comprehensive and reliable estimate of model generalization compared with a single held-out split. The results are summarized in Table~\ref{tab_res_cv}.

In the CAPE-V prediction task, VOQANet+ consistently outperforms VOQANet at both the utterance and patient levels across PVQD-A and PVQD-S subsets. For example, on PVQD-A (utterance level), VOQANet+ reduces RMSE from 9.656~$\pm$~0.648 to 9.376~$\pm$~0.228 and improves PCC from 0.847~$\pm$~0.046 to 0.870~$\pm$~0.049. Similarly, on PVQD-S (patient level), RMSE decreases from 7.343~$\pm$~0.491 to 6.266~$\pm$~0.843 and PCC increases from 0.919~$\pm$~0.018 to 0.958~$\pm$~0.018. 

Furthermore, in the GRBAS prediction task on the PVQD-S subset, a similar trend is observed. VOQANet+ achieves lower RMSE (0.324~$\pm$~0.016 compared with 0.334~$\pm$~0.023) and higher PCC (0.868~$\pm$~0.026 compared with 0.857~$\pm$~0.035) at the utterance level, corresponding to relative improvements of 3.0\% and 1.3\%, respectively. At the patient level, VOQANet+ also achieves lower RMSE (0.289~$\pm$~0.015 compared with 0.299~$\pm$~0.022; a 3.3\% reduction) and higher PCC (0.900~$\pm$~0.018 compared with 0.889~$\pm$~0.028; a 1.2\% improvement). These consistent patterns across all evaluation conditions further confirm that the integration of JSH features enhances both predictive accuracy and model stability, demonstrating the robustness of VOQANet+ in estimating perceptual voice quality across distinct rating protocols. These consistent improvements indicate that integrating JSH features enhances the model’s sensitivity to fine-grained acoustic cues associated with perceptual voice quality.

\subsection{Robustness to Seen and Unseen Noise}
To examine the generalizability of the models under adverse acoustic conditions, we tested VOQANet and VOQANet+ under various noisy conditions. Prior studies emphasize the need for noise-robust acoustic representations in dysphonic voice modeling~\cite{Zhang2022,Kuo2023}, highlighting domain robustness as essential for clinical use. As shown in Table~\ref{tab_robustness_test}, both models were evaluated for seen (i.e., white, pink, babble, and cocktail party noise used during training) and unseen (i.e., baby cry, laughter, and brown noise not included in training) noise types at multiple SNR levels from -5 dB to 10 dB. On both CAPE-V prediction and GRBAS prediction tasks, VOQANet+ performs slightly better than VOQANet under noisy conditions, demonstrating a consistent trend of improved stability across both seen and unseen conditions. For patient-level GRBAS prediction on PVQD-S under unseen noise, VOQANet+ improves PCC from 0.832 to 0.855 and reduces RMSE from 0.336 to 0.313. Similarly, for utterance-level CAPE-V prediction, PCC increases from 0.809 to 0.811 and reduces RMSE from 10.627 to 10.579. While numerically modest, these improvements represent a consistent performance trend under noisy clinical or telehealth scenarios, where maintaining reliable scoring in real-world acoustic conditions is important.

These results suggest that the inclusion of LLDs (JSH), which are rooted in perturbation measures and periodicity detection, may provide complementary information that helps stabilize performance under noisy conditions. By capturing signal-level voice irregularities that are partly independent of broadband spectral masking or background interference, VOQANet+ appears to benefit from representations that are somewhat less sensitive to additive noise. Furthermore, VOQANet+ also exhibits slightly smaller degradation from seen to unseen noise, particularly for GRBAS prediction, suggesting that combining interpretable acoustic features, clinically meaningful features with SFM embeddings can yield a more adaptable model that maintains consistent prediction quality across varied acoustic environments. These results support the potential of VOQANet+ for real-world clinical deployment, while acknowledging that further large-scale or cross-dataset evaluations would be valuable to substantiate this observation.

\subsection{Ablation Study on Different LLD Types (CPP vs. JSH)}
To further examine the influence of different LLDs on perceptual VQA, we compared cepstral-based and perturbation-based measures within the VOQANet+ framework. Previous clinical studies recommend cepstral-based measures, such as the Cepstral Peak Prominence (CPP), as reliable indicators of dysphonia across all phonation types (1–4)~\cite{Patel2018}, whereas perturbation measures like JSH are most valid for near-periodic Type 1–2 phonations.

We investigate two configurations for each feature type. The first is a configuration that uses only LLD (CPP or JSH) without SFM embeddings, and the second is the VOQANet+ configuration, which combines each LLD with WavLM (WS) embeddings. As shown in Table~\ref{tab:res_cpp}, the JSH configuration yields slightly lower RMSE and higher PCC than CPP on the sustained vowels, while CPP is more stable on connected speech. When combined with WavLM(WS), both feature types improve substantially, with JSH+WavLM(WS) providing the best overall results. For CAPE-V, JSH+WavLM(WS) achieves RMSE/PCC of 9.042/0.878 (PVQD-A) and 7.356/0.908 (PVQD-S), outperforming CPP+WavLM(WS) with 9.506/0.868 and 8.258/0.897. For GRBAS, JSH+WavLM(WS) also attains the highest PCC and lowest RMSE on both subsets. This finding aligns with our signal-type analysis, which confirmed that nearly all PVQD recordings correspond to Type 1–2 phonations with stable F$_0$ trajectories and distinct harmonic structures, while no highly aperiodic (Type 3–4) signals were observed, meaning that the perturbation-based features used in this study are acoustically valid for the dataset and provide complementary information to high-level SFM representations. Although cepstral measures such as CPP are theoretically more robust for severely irregular or noisy phonations, both feature types performed comparably in this work, with perturbation-based measures showing a slight advantage under the present recording conditions. 
It should be noted that cepstral measures remain the preferred acoustic indicators for highly irregular phonation (Types~3–4), while perturbation-based features are unreliable in this case.

Overall, these findings highlight a consistent trend across feature configurations. JSH alone performs slightly better than CPP for sustained vowels (PVQD-A), whereas CPP remains more stable for connected speech (PVQD-S). This result is consistent with acoustic principles because JSH depends on stable periodic phonation that is well represented in vowels but less reliable in connected speech with irregular voicing. When combined with SFM embedding, the JSH+WavLM(WS) configuration achieves better overall performance. That is because JSH captures small cycle-to-cycle variations that make the model more responsive to irregularities in voice quality, while CPP provides information that overlaps with the spectral content already captured by WavLM. 
We also acknowledge that dysphonic voices often exhibit short-term instability and intensity fluctuations, which may lead to the residual variability observed in acoustic analysis and model predictions.

These findings indicate that cepstral and perturbation measures capture complementary aspects of voice quality, where CPP reflects global harmonic organization and JSH captures subtle cycle-to-cycle variations in near-periodic phonations. Future research may explore adaptive feature selection based on estimated phonation type to improve cross-dataset generalizability. Furthermore, other complementary acoustic measures can be integrated, such as nonlinear dynamic or correlation-dimension analyses corresponding to perceptual ratings and laryngeal status.

\subsection{VOQANet and VOQANet+ Tested on the Saarbrücken Voice Database}
The Saarbrücken Voice Database (SVD)~\cite{putzer1997,putzer2007} contains German voice recordings from 2,043 speakers, including 687 healthy and 1,356 pathological individuals covering 71 different voice disorders. Each recording session consists of four speaking tasks: one connected-speech sentence and three sustained vowels. Specifically, the connected-speech task is the pronunciation of the German sentence ``Guten Morgen, wie geht es Ihnen?'' (``Good morning, how are you?''), while the sustained phonations include vowels /a/, /i/, and /u/, each produced at four pitch types (low, normal, high, and low–high–low). This design provides a rich combination of glottal and prosodic variations suitable for both sustained-vowel and connected-speech analyses. The recordings were collected under controlled conditions, and the dataset has been widely used for studies on pathological voice detection and voice quality assessment.

To further examine the generalizability of our proposed framework, we evaluated VOQANet and VOQANet+ on SVD. The evaluation followed a binary classification protocol (Healthy vs. Pathological) using patient-disjoint splits to prevent speaker overlap between training and testing sets. For each speaker, multiple phonation types were available. Moreover, to align with the PVQD setup, we considered two representative subsets: SVD-A (sustained vowels /a/ at normal pitch) and SVD-S (connected-speech sentences). The final classification decision for each speaker was obtained by averaging posterior probabilities across all utterances from the same individual. Model performance was assessed in terms of Accuracy, F1-Score, and AUC, with mean and standard deviation reported over five folds.

Table~\ref{tab:res_svd} summarizes the results. VOQANet+ (WavLM (WS) + JSH) achieved consistently higher metrics than VOQANet (WavLM (WS)) across both subsets, with average gains of approximately 3--10\% in Accuracy and F1-Score, and 2--4\% in AUC. Although the absolute differences are moderate, the trend was consistent across all folds, suggesting that incorporating perturbation-based low-level descriptors provides complementary cues that help stabilize predictions across datasets with different recording conditions, speakers, and languages. These findings indicate that the proposed approach can generalize reasonably well beyond the training corpus, even without dataset-specific fine-tuning. However, we acknowledge that further validation on additional corpora and under more diverse clinical conditions would be necessary to confirm its robustness and cross-lingual applicability. Overall, the SVD evaluation supports the potential of VOQANet+ as a flexible framework for pathological voice assessment across both sustained-vowel and continuous-speech tasks.

\section{Conclusions}
This study introduces VOQANet, a deep learning-based framework with an attention mechanism for automated pathological voice quality assessment. VOQANet uses SFM to extract representations from speech input, effectively capturing high-level acoustic and prosodic information from raw waveforms. It achieves strong predictive performance on both CAPE-V and GRBAS rating scales, demonstrating the utility of SFM embeddings in modeling perceptual voice characteristics of disordered voice.

By combining the strengths of pre-trained SFM representations and clinically interpretable acoustic features, VOQANet+ provides a robust and interpretable foundation for mildly dysphonic voices in real-world scenarios. 
Although the current study relied on CAPE-V annotations provided in the PVQD dataset,  the proposed framework is fully compatible with CAPE-Vr, which offers updated guidelines for perceptual assessment. Thus, future dataset adopting CAPE-Vr can be readily integrated, further aligning automated VQA with current clinical standards.
While the current study focused on sustained vowels and short CAPE-V sentences, these tasks provide limited coverage of prosodic variation such as intonation and rhythm. Furthermore, it could incorporate longer or context-rich speech passages to capture prosodic cues that may further enhance pathological voice assessment. 
Future research directions may also include exploring multi-task learning to jointly predict individual CAPE-V (or GRBAS) dimensions or perceptual subscales, as well as cross-lingual generalization and domain adaptation to enable broader deployment across clinical settings and languages.

\end{document}